\documentclass[12pt]{article}

\usepackage{amsmath}
\usepackage{amssymb}
\usepackage{epsfig}

\newcommand{\capdef}{}
\newcommand{\mycaption}[2][\capdef]{\renewcommand{\capdef}{#2}%
        \caption[#1]{{\itshape #2}}} 
\makeatletter
\renewcommand{\fnum@table}{\textbf{\tablename~\thetable}}
\renewcommand{\fnum@figure}{\textbf{\figurename~\thefigure}}
\makeatother

\newcounter{myenumi}

\renewcommand{\themyenumi}{\roman{myenumi}}
{\end{list}}

\newlength{\myem}
\newcommand{\sep}[1]{#1}
\newcounter{mysubequation}[equation]
\renewcommand{\themysubequation}{\alph{mysubequation}}
\newcommand{\mytag}{\stepcounter{mysubequation}%
\tag{\theequation\protect\sep{\themysubequation}}}
\newcommand{\globallabel}[1]{\refstepcounter{equation}\label{#1}}

\makeatletter
\renewcommand{\section}{\@startsection{section}{1}{0em}{-\baselineskip}%
{\baselineskip}{\normalfont\large\bfseries}}
\renewcommand{\subsection}%
{\@startsection{subsection}{2}{0em}{-0.7\baselineskip}%
{0.7\baselineskip}{\normalfont\bfseries}}
\makeatother

\newcommand{\PCPV}{
\begin{picture}(22,10)
\put(8,-2){\line(2,1){12}}
\put(0,0){$P_{CP}$}
\end{picture}}

\newcommand{\PCPC}{
\begin{picture}(22,10)
\put(0,0){$P_{CP}$}
\end{picture}}

\newcommand{\CPC}{{\text{CP}}}

\newcommand{\pdagger}{{\phantom{\dagger}}}

\newcommand{\GeV}{\,\mathrm{GeV}}
\newcommand{\MeV}{\,\mathrm{MeV}}

\newcommand{\eV}{\,\mathrm{eV}}

\newcommand{\fracwithdelims}[4]{\left#1 \frac{#3}{#4} \right#2}
\newcommand{\vev}[1]{\left\langle #1\right\rangle}

\newcommand{\Ahalf}[0]{\frac{1}{2}}

\DeclareMathOperator{\im}{Im}
\DeclareMathOperator{\re}{Re}

\DeclareMathOperator{\diag}{Diag}

\begin{document}


\begin{titlepage}

\renewcommand{\thefootnote}{\alph{footnote}}

\begin{flushright}
  {\hfill TUM--HEP--345/99}\\
  {SFB 375--334}\\
  {MPI--PhT/9907}\\
  {OUTP--99--15P}\\
  {April 1999}
\end{flushright}

\vspace*{0.5cm}

\renewcommand{\thefootnote}{\fnsymbol{footnote}}
\setcounter{footnote}{-1}

{\begin{center} {\LARGE\bf CP-Violation in Neutrino
      Oscillations$^*$\footnote{\hspace*{-1.6mm}$^*$Work supported in part by
        "Sonderforschungsbereich 375 f\"ur Astro-Teilchenphysik" der Deutschen
        Forschungsgemeinschaft and by the TMR Network under the EEC
        Contract No.~ERBFMRX--CT960090.}}
\end{center}}
\renewcommand{\thefootnote}{\alph{footnote}}

\vspace*{.8cm}
{\begin{center} {\large{\sc 
                K.~Dick\footnote[1]{\makebox[1.cm]{Email:}
                Karin.Dick@physik.tu-muenchen.de}\footnotemark[2],
                M.~Freund\footnote[3]{\makebox[1.cm]{Email:}
                Martin.Freund@physik.tu-muenchen.de},  
                M.~Lindner\footnote[4]{\makebox[1.cm]{Email:}
                Manfred.Lindner@physik.tu--muenchen.de}
                \vspace{0.2cm}
                and
                {\sc 
                A.~Romanino\footnote[5]{\makebox[1.cm]{Email:}
                romanino@thphys.ox.ac.uk}}}}
\end{center}}
\vspace*{0cm}
{\it 
\begin{center}  

   \footnotemark[1]${}^,$\footnotemark[3]${}^,$\footnotemark[4]%
                Institut f\"ur Theoretische Physik,
                Technische Universit\"at M\"unchen,\\
                James--Franck--Strasse, D--85748 Garching, Germany
   \vskip .3cm

   \footnotemark[2]%
                Max-Planck-Institut f\"ur Physik,
                Postfach 401212, D--80805 M\"unchen, Germany
   \vskip .3cm

   \footnotemark[5]%
                Department of Physics, Theoretical Physics, University of
                Oxford,\\ Oxford OX13NP, UK

\end{center} }
\vspace*{0.5cm}
{\Large \bf 
\begin{center} Abstract \end{center}  }
We study in a quantitative way CP-violating effects in neutrino oscillation
experiments in the light of current and future data. Different scenarios with
three and four neutrinos are worked out in detail including matter effects 
in long baseline experiments and it is shown that in some
cases CP-violating effects could affect the analysis of a possible measurement.
In particular in the three neutrino case we find that the effects can be larger
than expected, at least in long-baseline $\nu_\mu\rightarrow\nu_e$. Moreover,
measuring these effects could give useful information on the solar oscillation
frequency. In four neutrino scenarios large effects are possible both in the
$\nu_\mu\rightarrow\nu_\tau$ and $\nu_\mu\rightarrow\nu_e$ channels of
long-baseline experiments, whereas short-baseline experiments are affected only
marginally.

\end{titlepage}

\newpage

\renewcommand{\thefootnote}{\arabic{footnote}}
\setcounter{footnote}{0}


\section{Introduction \label{sec:intro}}

Understanding the mechanism responsible for the patterns and values 
of fermion masses, mixings and CP-phases is a very important goal 
of particle physics. The recent advances in the field of neutrino 
physics, especially by experiments measuring neutrino oscillations, 
are in this context extremely valuable. There is now overwhelming 
evidence for neutrino masses and it is possible to extract from 
data very interesting patterns for masses, mixings and as we will
see CP-phases. Neutrino masses require extra ingredients, which 
constitute a small extension of the Standard Model and could be 
viewed as a complication, but it seems likely that this new 
information is an important second lever arm for the fermion mass 
problem. The reason is that in many extensions of the Standard Model, 
like for example in GUT theories, the apparent similarities between 
quarks and leptons lead to connections between lepton and quark 
masses via the same symmetry breaking Higgses and/or related 
Yukawa couplings. The new results from the neutrino sector can 
therefore often be combined with information from the quark sector 
allowing thus a better test of proposed mass structures (like e.g. 
in the form of so--called textures). Neutrino masses imply mixings 
in the lepton sector which include also CP-violating phases. 
These CP-phases are however not only interesting due to their 
appearance in mass textures, but they can be responsible for very 
important physical effects. Mechanisms have for example been proposed, 
where CP-violation in the neutrino sector is the essential ingredient 
in explaining the baryon asymmetry of the universe. CP-violation 
in the neutrino sector is thus by itself an important issue.

Besides these questions of general theoretical kind the presence of 
CP-violation can also have quite significant impact on how experiments
should be performed and analyzed. We will discuss the possibility of 
having CP-violation in various experiments from a qualitative and 
quantitative point of view and will point out that the effects can be 
larger than expected from qualitative arguments~\cite{others}.  
In long baseline $\nu_\mu\rightarrow\nu_e$ experiments we will find 
for example with three (four) neutrinos and maximal CP-phase 
CP-asymmetries up to 40\% (60\%), while short baseline experiments 
could see in some corners of parameter space of $\nu_\mu$--$\nu_\tau$ 
oscillations effects up to 10\% with four neutrinos, so that a modest 
improvement of the baseline would be enough to see sizeable effects. 
CP-violation can thus affect in a significant 
way (or even spoil) the measured probabilities, such that the planning of
experiments, data analysis and theoretical interpretation should take 
them into account. Furthermore measuring CP-violating effects would 
provide information on the frequency giving rise to the solar oscillation, 
at least in the framework with $\nu_\mu\rightarrow\nu_e$ transition and 
three neutrinos. In this respect CP-violation offers an interesting 
possibility to investigate in long-baseline experiments parameters 
typically accessible through solar neutrino experiments. This is because 
the solar frequency suppression of CP-violation is only linear and 
because a $\theta_{13}$ angle suppression affect the CP-conserving 
amplitude more than the CP-violating one. 

The minimally extended Standard Model with three heavy right-handed neutrinos
allows usually only for two independent oscillation frequencies, while
experiments claim three different values. Therefore even complex three 
neutrino scenarios, where more than one pair of neutrinos oscillates, 
can not accommodate all three results simultaneously. Once this possibility 
is excluded one 
can either extend the theoretical framework such that all data can be 
accommodated or one must discuss different scenarios excluding some of
the data. We will consider both possibilities. Since the LSND evidence is
considered most controversial we will exclude their result in an analysis 
with three neutrinos. The analysis of four neutrino scenarios is done 
with all available experimental information including LSND. 
In this case a third squared mass difference $\Delta m^2_{\text{LSND}}$ much
larger than the solar and atmospheric ones explains the LSND oscillation 
signal and can generate small CP-violation effects even in short-baseline 
experiments sensitive to this squared mass difference.

Our paper is organized as follows. First we introduce our notation and
framework. Next we give a general discussion of potential CP-violating 
effects in neutrino oscillation experiments and present the existing 
experimental data that we use in our analysis. Then follows an extensive
analysis of CP-violating effects in scenarios involving three or four 
neutrino species. In the end all results are discussed in comparison and 
conclusions for planned neutrino oscillation experiments are drawn.


\section{CP-violation in neutrino 
         oscillation experiments \label{sec:CPviolation}} 

\subsection{Masses, mixings and notation}

The three neutrino flavour eigenstates $\nu_{e_i}$, $i\in\{1,2,3\}$ 
form in the Standard Model (SM) with their respective charged lepton 
partners doublets under $SU(2)_L$ and without loss of generality we 
can define the neutrino flavour eigenstates in a basis where the 
charged lepton mass matrix is already diagonal. Neutrino oscillations 
require non-degenerate (Dirac or Majorana) neutrino masses, which are 
however forbidden in the SM, since any mass term or Yukawa coupling 
which generates neutrino masses would violate the gauge symmetry. 
Neutrino oscillations require therefore SM extensions with new fields 
and/or interactions. A new $SU(2)_L$ Higgs triplet field, for example, 
with hypercharge $Y=2$ and non-vanishing vacuum expectation value can 
generate a $3\times 3$ Majorana mass matrix $M_L$ for the neutrino 
fields $\nu_{e_i}$ via Yukawa
couplings\footnote{Note that for phenomenological reasons the VEV of such a
  triplet should be rather small compared to the electro-weak scale since it
  would contribute already at tree level to custodial SU(2) breaking.
  Furthermore the masses of the so far unobserved single and double charged
  Higgses should be very large.}.  
Another possibility is the existence of further neutral and 
sterile\footnote{I.e. $SU(2)_L$ singlets. The existence of
  further non-singlet $SU(2)_L$ representations containing a neutral fermion
  is strongly disfavoured. Such representations would require further fermions
  to satisfy the anomaly conditions. Existing mass bounds for new fermions
  (e.g.  generations) would then lead to big unobserved radiative corrections
  in the so--called S and T variables.}  
neutrino states $\nu_{e_j}$, $j\in\{4,\ldots, n_N\}$ resulting in the most 
general mass term in the Lagrangian
\begin{equation}
{\cal L}_{mass}= (m_N)_{ij}~{\nu}_{e_i}{\nu}_{e_j} + h.c.~,
\label{Lmass}
\end{equation}
with the extended $n_N\times n_N$ neutrino mass matrix
\begin{equation}
m_N=
\begin{pmatrix}
  \Ahalf M_{L} & M_D \\
  M_D^T & \Ahalf M_{R}
\end{pmatrix}~.
\label{Mmass}
\end{equation}
In addition to $M_L$ this symmetric mass matrix contains the $3\times
(n_N-3)$ dimensional off-diagonal entries $M_D$ which can be generated by
Dirac--like Yukawa couplings and the electro-weak VEV.  Note that the
remaining $(n_N-3) \times (n_N-3)$ dimensional sub-mass matrix $M_R$ can 
be made of explicit mass terms with arbitrary values, 
since none of the symmetries of
the Lagrangian is in this way violated.  The elements of $M_R$ are thus
uncorrelated with the electro-weak sector and a natural range for these mass
terms is the scale where the new fields become members of multiplets in some
extended framework\footnote{For $n_N=6$ the three new neutrino fields can for
  example be placed very economically together with $e_{\nu i}$,
  $i\in\{4,5,6\}$ in doublets of $SU(2)_R$ in left--right symmetric models.
  Contributions to $M_R$ arise then via left--right symmetric Yukawa
  interactions resulting in masses of the order of the left--right breaking
  scale. Alternatively the new neutrino fields can be fitted into
  representations of some GUT group and natural entries in $M_R$ would in this
  case be at the GUT scale.}.
This implies that some (but not necessarily all) eigenvalues of $M_R$ are 
heavy enough to decouple after the diagonalization of the matrix in 
eq.~(\ref{Mmass}), thus leaving only $n\geq 3$ light states involved in the
low energy physics\footnote{Another reason why some neutrino states might
decouple is that there are two degenerate eigenstates one of which can be made
decoupled with a rotation in their subspace. This would be the case for example
if only the non-diagonal blocks in eq.~(\ref{Mmass}) were non-vanishing 
(pure Dirac case).}.

Independently of the physics giving rise to them, we will discuss in this paper
the case of three or four light neutrino degrees of freedom, i.e.\ $n=3,4$.
These mass eigenstates are assumed to be mixtures of the original three flavour
eigenstates which couple to $SU(2)_L$ and possibly with other sterile states
$\nu_{e_i}$, $i\in\{4,\ldots, n\}$. In the limit where all heavy neutrino
degrees of freedom are decoupled we can thus write down a $n\times n$ neutrino
mass matrix which leads upon diagonalization to $n=3,4$ mass eigenstates and a
$n\times n$ CKM matrix $U_{CKM}$.  Neutrino oscillation is in this picture to a
very good approximation the oscillation of $n$ neutrino degrees of freedom,
where only one of the first three flavour eigenstates can be produced and
detected (via its coupling to $W's$). These flavour eigenstate can be written
with the help of $U_{CKM}$ as a superposition of the $n$ mass eigenstates. The
transition probability becomes then
\begin{equation}
P(\nu_{e_i}\rightarrow\nu_{e_j})
= \left | \left( U_{CKM} D U_{CKM}^+ \right)_{ij} \right |^2 ~,
\end{equation}
where $n\times n$ matrix multiplication is implied, 
$D= \diag(e^{-iE_1t}, \dots, e^{-iE_nt})$
and where the indices $i,j$ correspond
to the respective flavour eigenstate with $i,j\in\{1,2,3\}$.

The $n\times n$ CKM matrix $U_{CKM}$ contains as usual a number of 
global unphysical phases which can be absorbed into the fermion fields. 
Because of the potential Majorana nature of the neutrinos more physical 
phases survive, however, compared to the quark case with pure Dirac 
masses. Altogether there are up to $n(n-1)/2$ physical mixing angles and 
up to $n(n-1)/2$ physical phases. For $n=3$ ($n=4$) we have thus when all
masses are non-degenerate and not purely Dirac-like three (six) 
mixing angles and three (six) physical phases. In the $n=3$ case we 
use the standard parameterization
\begin{equation} 
U_{CKM} = 
U(\theta_{12},\theta_{23},\theta_{13},\delta)
\diag(e^{i\alpha_1},e^{i\alpha_2},1)~, 
\label{Ual} 
\end{equation} 
where the so--called CKM-like phase $\delta$ and the extra Majorana 
phases $\alpha_1$ and $\alpha_2$ are explicitely shown. With the help of  
$c_{ij}=\cos(\theta_{ij})$ and $s_{ij}=\sin(\theta_{ij})$ 
we can chose as usual:
\begin{equation} 
U(\theta_{12},\theta_{23},\theta_{13},\delta)= 
\begin{pmatrix} 
  c_{12}c_{13} & s_{12}c_{13} & s_{13}e^{-i\delta} \\
  -s_{12}c_{23} -c_{12}s_{23}s_{13}e^{i\delta} & 
  c_{12}c_{23}
  -s_{12}s_{23}s_{13}e^{i\delta} & s_{23}c_{13} \\
  s_{12}s_{23} -c_{12}c_{23}s_{13}e^{i\delta} & 
  -c_{12}s_{23}
  -s_{12}c_{23}s_{13}e^{i\delta} & c_{23}c_{13}
\end{pmatrix}. 
\label{Ut1} 
\end{equation} 
For $n>3$ we will not use any particular parameterization, but we can still 
factorize $U_{CKM}=U\cdot \diag(e^{i\alpha_1},\dots,e^{i\alpha_{n-1}},1)$. 
Since only $U_{CKM} D U_{CKM}^+$ is involved in oscillation experiments, 
and since $D$ is diagonal and commutes with the extra diagonal Majorana 
phases $\diag(e^{i\alpha_1},\dots,e^{i\alpha_{n-1}},1)$, it is always 
possible to eliminate those extra phases. Thus only $(n^2-3n+2)/2$ phases 
(one for $n=3$, three for $n=4$) show up in oscillation experiments
just like in the quark case. Due to this similarity we call these 
remaining phases ``CKM-like''. 

\subsection{CP-violation} 

The oscillation probability for a neutrino produced in a flavour eigenstate
$\nu_{e_i}$  to be detected as a $\nu_{e_j}$ after having traveled a 
distance $L$ with a ultra-relativistic energy $E$ is
\globallabel{Pne}
\begin{align}
P(\nu_{e_i}\rightarrow\nu_{e_j}) &=
\PCPC(\nu_{e_i}\rightarrow\nu_{e_j}) + \PCPV(\nu_{e_i}\rightarrow\nu_{e_j})~,
\mytag \\
P(\bar{\nu}_{e_i}\rightarrow\bar{\nu}_{e_j}) &=
\PCPC(\nu_{e_i}\rightarrow\nu_{e_j}) - \PCPV(\nu_{e_i}\rightarrow\nu_{e_j})~,
\mytag
\end{align} 
where
\globallabel{PCP}
\begin{align}
\PCPC(\nu_{e_i}\rightarrow\nu_{e_j}) &= \delta_{ij}
-4\sum_{k>h}\re(J^{e_j e_i}_{kh})\sin^2\Delta_{kh}~, \mytag \\
\PCPV(\nu_{e_i}\rightarrow\nu_{e_j}) &= 4\sum_{k>h}\im(J^{e_j
e_i}_{kh})\sin\Delta_{kh}\cos\Delta_{kh}~, \mytag
\end{align} 
with $J^{e_i e_j}_{kh} = U_{e_i\nu_k} U_{\nu_k e_j}^\dagger
U_{e_j\nu_h} U_{\nu_h e_i}^\dagger$, $J^{e_i e_j}_{kh}=(J^{e_j
e_i}_{kh})^*=(J^{e_i e_j}_{hk})^*$ and $\Delta_{kh}=\Delta m^2_{kh}
L/4E$. Notice also the relations $\sum_k J^{e_i e_j}_{kh} =
\delta_{ij} |U_{ih}|^2$, $\sum_i J^{e_i e_j}_{kh} =
\delta_{kh} |U_{jh}|^2$. One can easily see that only $n-1$ out 
of the $n(n-1)/2$ frequencies $|\Delta m^2_{kh}|$, $k>h$, are independent.  
From CPT invariance follows in general $P(\nu_{e_i}\rightarrow\nu_{e_j}) =
P(\bar{\nu}_{e_j}\rightarrow\bar{\nu}_{e_i})$ and in particular
$P(\nu_{e_i}\rightarrow\nu_{e_i}) =
P(\bar{\nu}_{e_i}\rightarrow\bar{\nu}_{e_i})$.
Therefore one can see that CP-violating effects can not occur in 
disappearance experiments.

Finding sizeable CP-violating effects in oscillations is however not easy 
since they are affected by suppressions similar to the quark case. First 
$|\PCPV|\leq \PCPC$, since $\PCPC+\PCPV$ and $\PCPC-\PCPV$ have
both to be positive. Moreover, the CP-violating contribution to the 
oscillation probability is suppressed, because CKM-like CP-violation 
is not possible with only two neutrinos (or only two non-degenerate 
neutrinos).  
As a consequence CP-violating effects need (at least) three
mixing angles between non-degenerate neutrinos such that the squared
mass difference corresponds to wavelenghts neither too large compared
with the distance travelled (otherwise the oscillation does not have
enough time to develop) nor too short (otherwise the effect is washed out).
These requirements are made explicit by the three neutrino version of
eqs.~(\ref{Pne},\ref{PCP}). Using $\sigma_{ij}\equiv \sum_k \varepsilon_{ijk}$ 
and $\im J^{e_i e_j}_{kh} = -\sigma_{ij}\sigma_{kh}J_{\CPC}$
it is in fact in this case
\globallabel{3n}
\begin{align}
\PCPC(\nu_{e_i}\rightarrow\nu_{e_j}) &= \delta_{ij} -4\re J^{ji}_{12}
\sin^2\Delta_{12} -4\re J^{ji}_{23} \sin^2\Delta_{23} -4\re J^{ji}_{31} 
\sin^2\Delta_{31}~, \mytag \\
\PCPV(\nu_{e_i}\rightarrow\nu_{e_j}) &= -8\sigma_{ij} J \sin\Delta_{12} 
\sin\Delta_{23} \sin\Delta_{31}  \nonumber \\
& = 8\sigma_{ij}\, J(\sin^2\Delta_{23}\sin\Delta_{12}\cos\Delta_{12} 
+\sin^2\Delta_{12} \sin\Delta_{23} \cos\Delta_{23})~, \mytag  
\end{align}
with
\begin{equation}
8 J = \cos\theta_{13} \sin(2\theta_{13}) \sin(2\theta_{12})
\sin(2\theta_{23}) \sin\delta~.
\label{Jct}
\end{equation}
From eqs.~(\ref{3n}\sep{b},\ref{Jct}) we see explicitely that even in
case of maximal CP-violation (i.e. $|\sin\delta|=1$) small angles and 
small $\sin\Delta_{12}$ suppress the effect. On the other hand it is 
important to notice that these two suppressions are only linear so that 
in cases in which it is safe to neglect $\sin^2\Delta_{12}$ effects in 
the CP-conserving part of the probability, still CP-violating effects 
proportional to $\sin\Delta_{12}$ can be relevant. Experiments sensitive 
to CP-conserving effects with amplitudes suppressed by the $\sin^2$ of 
a small angle, can thus have a chance to see a CP-violation effect only 
proportional to the $\sin$ of that angle. We will study this in a 
quantitative way in the following.

\subsection{Asymmetries}\label{subsec:asymmetries}

As a measure of CP-violation, we will consider suitable asymmetries 
between CP-conjugated transitions. Besides having obvious physical 
meaning these asymmetries will show to what extent the analysis of 
a possible signal in a single channel $\nu_{e_i}\rightarrow\nu_{e_j}$ 
(or $\bar{\nu}_{e_i}\rightarrow\bar{\nu}_{e_j}$) performed without 
taking into account CP-violation could be spoiled by CP-violation 
effects. Neutrinos travel in some experiments through matter such 
that the two conjugate channels have to be distinguished in the 
vicinity of the corresponding MSW region. Where appropriate we will 
use therefore the suffix ``${\text{m}}$'' to express that
the transition probabilities in eqs.~(\ref{Pne}) are changed by the 
presence of matter. With this in mind we define the total asymmetry
\begin{equation}
a^{\text{tot}}_{ij} = \frac{ \vev{P^{\text{m}}(\nu_{e_i}\rightarrow\nu_{e_j})}
-\vev{P^{\text{m}}(\bar{\nu}_{e_i}\rightarrow\bar{\nu}_{e_j})} }{
\vev{P^{\text{m}}(\nu_{e_i}\rightarrow\nu_{e_j})} +
\vev{P^{\text{m}}(\bar{\nu}_{e_i}\rightarrow\bar{\nu}_{e_j})} }~,
\label{aij}
\end{equation}
where the average symbol $\langle\rangle$ in eq.~(\ref{aij}) accounts for 
the averaging in energy and length present in every real experiment and 
is particularly important in case both the channels, and therefore the 
asymmetry itself, are measured. In this case
\globallabel{ave}
\begin{align}
\vev{P^{\text{m}}(\nu_{e_i}\rightarrow\nu_{e_j})} &= \int d(L/E)
P^{\text{m}}{(\nu_{e_i}\rightarrow\nu_{e_j})}(L/E) f(L/E) ~, \mytag \\
\vev{P^{\text{m}}(\bar{\nu}_{e_i}\rightarrow\bar{\nu}_{e_j})} &= \int d(L/E)
P^{\text{m}}{(\bar{\nu}_{e_i}\rightarrow\bar{\nu}_{e_j})}(L/E) \bar{f}(L/E)~,
\mytag 
\end{align}
where the weight functions $f$, $\bar{f}$ include the initial spectra, the
cross-sections, the efficiencies and the resolutions and can be assumed to 
be normalized to 1 without loss of generality but in general do not have 
the same shape\footnote{The original weight function $f_0(L,E)$ is given 
by the experimental energy and length spectrum and depends thus on $L$ 
and $E$. Since $P$ is a function of $L/E$ only we can introduce the length 
averaged weight function $f(L/E)=\int{(E^2/L)~f_0(L,E/L\cdot L)}dL$.}. 
The total asymmetry $a^{\text{tot}}_{ij}$ receives three different 
contributions. The first comes from the weight functions $f$, 
$\bar{f}$ and $f\neq \bar f$ corresponds to an asymmetry in the 
experimental apparatus. If matter effects are relevant then 
there is a second experimental asymmetry to be distinguished 
from the third ``intrinsic'' asymmetry due to CP-violation. 
Matter effects and $f\neq\bar{f}$, i.e. the CP-violation of 
the experimental setup, give thus rise to a non-vanishing asymmetry 
$a^{\text{exp}}_{ij}$ even when the mixing matrix $U$ is real.
It is thus very important to distinguish the experimental asymmetry 
from the asymmetry $a_{ij}^\CPC$ due to intrinsic CP-violation
which can be written as~\cite{gavela98}
\begin{equation}
a_{ij}^\CPC \equiv  a_{ij}^{\text{tot}} - a_{ij}^{\text{exp}}~.
\label{Dasy}
\end{equation}
By definition, $a_{ij}^\CPC$ must vanish upon setting all CP-violating 
phases to zero and measuring a non-vanishing $a_{ij}^\CPC$ would be
a signal of intrinsic CP-violation in the vacuum mixing matrix $U$. 
Extraction of $a_{ij}^\CPC$ from data requires of course knowledge of 
$a^{\text{exp}}_{ij}$ which is only possible if the CP-conserving 
parameters involved in neutrino oscillation are known. This is most
likely the case once a measurement of CP-violation in neutrino 
oscillation becomes feasible. The uncertainties of $a^{\text{exp}}_{ij}$ 
will however add to the uncertainties on $a_{ij}^\CPC$.

The dependence of the total asymmetry on the weight functions and
on matter effects is in general rather complicated. It is therefore 
not easy to relate the asymmetry $a_{ij}^\CPC$ of different 
experimental setups or to a reference setup with vanishing matter 
effects and CP-invariant apparatus, i.e. oscillation in vacuum and 
the same averaging function $f_+ = (f+\bar{f})/2$ for both channels.
For the scenarios considered in this paper we find however that matter
effects are small enough such that they can be considered as 
a correction to the case without matter. 

We omit therefore from now on the suffix ``${\text{m}}$'' and 
discuss $a_{ij}^\CPC$ in vacuum analytically. The numerically
calculated corrections due to matter effects will be presented 
afterwards. If $f_\pm = (f\pm\bar{f})/2$ are the average vacuum 
weight function and asymmetry, respectively, then one finds upon 
neglecting a twice suppressed 
$\int d(L/E) \PCPV{(\nu_{e_i}\rightarrow\nu_{e_j})}(L/E) f_-(L/E)$ 
in the denominator of $a_{ij}^{\text{tot}}$,
\begin{equation}
a_{ij}^{\text{tot}} \simeq a_{ij}^\CPC + a_{ij}^{\text{exp}} ~,
\label{2asy}
\end{equation}
where
\globallabel{exp0}
\begin{align}
a^{\text{exp}}_{ij} &= \frac{\int d(L/E)
\PCPC{(\nu_{e_i}\rightarrow\nu_{e_j})}(L/E) f_-(L/E)}{\int d(L/E)
\PCPC{(\nu_{e_i}\rightarrow\nu_{e_j})}(L/E) f_+(L/E)}~, \mytag \\
a^\CPC_{ij} &= \frac{\int d(L/E)
\PCPV{(\nu_{e_i}\rightarrow\nu_{e_j})}(L/E) f_+(L/E)}{\int d(L/E)
\PCPC{(\nu_{e_i}\rightarrow\nu_{e_j})}(L/E) f_+(L/E)}~. \mytag
\end{align}
The quantity $a^{\text{exp}}_{ij}$ is now the remaining contribution 
to $a_{ij}^{\text{tot}}$ coming from the experimental apparatus, 
or more precisely the asymmetry averaged with the CP-conserving 
part of the probability, whereas $a^\CPC_{ij}$ is the CP-violating 
contribution to $a_{ij}$, given by the CP-asymmetry averaged with $f_+$.
A nice feature of the asymmetry $a^\CPC_{ij}$ in vacuum is that it 
does not depend too much on the function $f_+$, namely on the details 
of the experimental apparatus. Most of this dependence cancels in fact 
in the ratio, especially when the two frequencies giving rise to the 
asymmetry are well separated. $|a^\CPC_{ij}|<1$ follows from 
$|\PCPV|<\PCPC$.  We will study in the following first 
the CP-violating contribution $a^\CPC_{ij}$ in vacuum, assuming that 
$a^{\text{exp}}_{ij}$ is small or under control. Matter corrections
will be presented subsequently.

\subsection{Existing experimental data \label{sec:data}}

Today, there are three different kinds of experiments, which are in favour of
neutrino oscillations. First there is the long known solar neutrino-deficit
which can be explained by an oscillation $\nu_e \rightarrow \nu_x$. In the
context of the standard solar model \cite{BAHSSM98}, the data, which are mainly
obtained from Super-Kamiokande, SAGE, Gallium and Chlorine experiments, can be
fitted by a two neutrino oscillation either in vacuum or matter enhanced
(MSW-effect). The vacuum solution requires a large mixing with 
$\Delta m^2 \approx (0.5 - 2.0)\cdot 10^{-9} \mathrm{eV}^2$. For the 
MSW-solution we have three different allowed parameter 
regions~\cite{bahcall:98a} where the low-$\Delta m^2$
solution is somewhat disfavoured compared to the solutions with high $\Delta
m^2$, which split into a small mixing (SMA) and large mixing (LMA) case.  The
SMA solution with $\Delta m^2\approx (0.4 - 1.0)\cdot 10^{-5}\mathrm{eV}^2$ 
and the LMA solution with $\Delta m^2 \approx (0.2 - 2.0)\cdot
10^{-4}\mathrm{eV}^2$ (at $99\%$ C.L.)  are obtained from global fits of the
rates measured by all solar experiments and the day/night measurements of
Super-Kamiokande~\cite{bahcall:98a,SK:98b}.  
The LMA solution\footnote{If the last data points of the electron recoil 
energy spectrum in Super-Kamiokande are reliable, then they would tend to 
favour the SMA solution and to disfavour the LMA solution. This is 
however still largely an open problem and we will delay this issue
therefore.}
will be especially interesting
in the case of three neutrinos from the point of CP-violating effects, while
scenarios with more neutrinos have interesting CP-violating effects even
without the LMA solution. For CP-violating effects in connection with the 
LMA-solution it is also important that the upper bound for $\Delta m^2$ 
depends especially on the Chlorine experiment~\cite{BHSSW}. As a consequence, 
excluding or weakening the Chlorine results in the analysis can significantly 
enlarge the possible CP-violating effects.

The second evidence for neutrino oscillations is given by the atmospheric 
neutrino data. The Super-Kamiokande experiment measures a zenith angle 
dependent flux of atmospheric muon neutrinos which can be explained by 
an oscillation of the type $\nu_{\mu} \rightarrow \nu_{\tau,s}$ with a 
value of $\Delta m^2 \approx (0.3 - 8)\cdot 10^{-3}\mathrm{eV}^2$ 
(at $99\%$ C.L.) and maximal mixing~\cite{SK98}. The ratio of neutrinos 
reaching the detector to the number of neutrinos produced in the atmosphere, 
$R_{\mu,e} = N(\nu_{\mu,e}+\bar{\nu}_{\mu,e}) /
N_0(\nu_{\mu,e}+\bar{\nu}_{\mu,e})$, is given by
\globallabel{RmR}
\begin{align}
R_\mu &= P(\nu_\mu\rightarrow\nu_\mu) +R^{-1}_0
\PCPC(\nu_e\rightarrow\nu_\mu) +R^{-1}_0 a_0^e
\PCPV(\nu_e\rightarrow\nu_\mu)~, \mytag \\
R_e &= P(\nu_e\rightarrow\nu_e) +R^{~}_0
\PCPC(\nu_e\rightarrow\nu_\mu) -R^{~}_0 a_0^\mu
\PCPV(\nu_e\rightarrow\nu_\mu)~, \mytag
\end{align}
with the particle-antiparticle asymmetry $a_0^{\mu,e}
= N_0(\nu_{\mu,e}-\bar{\nu}_{\mu,e}) / N_0(\nu_{\mu,e}+\bar{\nu}_{\mu,e})$ 
in the initial neutrino flux and the initial electron--muon asymmetry 
$R_{0} = N_{0}(\nu_\mu+\bar{\nu}_\mu) / N_{0}(\nu_e+\bar{\nu}_e)$. 
Super-Kamiokande measures a zenith angle dependence (and therefore a $L/E$ 
dependence) of $R_{\mu}$, but none for $R_e$.
 
A third indication for oscillation is claimed by the LSND experiment
\cite{LSND98}. It has to be remarked, however, that major parts of the LSND
allowed parameter region are in contradiction to the KARMEN experiment. This
results needs to be checked therefore by new experiments (e.g. MiniBooNE). 
If the LSND result is correct, then it implies oscillations of the
type $\bar{\nu}_{\mu} \rightarrow \bar{\nu}_e$ with a lower limit of
$10^{-1} \mathrm{eV}^2$ for the squared mass difference and good fits 
around one $\mathrm{eV}^2$.

Apart from these positive results, important constraints for the oscillation
parameters are provided by negative results of disappearance experiments of 
the type $\nu_e \rightarrow \nu_e$ and $\nu_{\mu} \rightarrow
\nu_{\mu}$~\cite{CHOOZ97,CHORUS98}. An important result stems from the Chooz
experiment, which allows to put an upper limit of about $10^{-3}\eV^2$ on all
frequencies contributing to $\bar{\nu}_e$ disappearance with an amplitude
larger than 0.2. The Bugey experiment found strong constraints~\cite{bugey} 
on the amplitudes of contributions to $\bar{\nu}_e$ disappearance with 
frequencies larger than $3\cdot~10^{-2}\eV^2$. On the other hand the 
CDHS and CCFR 
experiments~\cite{CDHS,CCFR} give a limit on the amplitudes of contributions 
to $\bar{\nu}_\mu$ disappearance with frequencies larger than $0.3 \eV^2$.
There exist also absolute neutrino mass limits, like for example from 
precise measurements of the endpoint in the $\beta$-decay spectrum,
which lead in principle to further upper bounds on $\Delta m^2$. 
These limits are currently however weaker than the ranges quoted above.


\section{CP-violation with three neutrinos \label{sec:3nu}}

As already mentioned it is not possible to accommodate the three 
different experimental signals for oscillation in scenarios with 
only three neutrinos, with three different squared mass differences, 
$|\Delta m^2_{ij}|$, $i<j$, among which only two are independent. 
One must therefore either exclude one evidence for oscillation from 
the analysis or postulate the existence of further neutrinos. Since 
these options are very different we will study here both possibilities.

The LSND evidence for oscillation is in a large part of the allowed parameter
space in contradiction with limits from KARMEN.  The LSND evidence for
oscillation is therefore by far the most controversial one, while the
atmospheric and solar ones seem more solid. We will therefore consider in this
section first the case in which there are only three light neutrinos and where
the LSND evidence is left out.  In the next section we study a four neutrino
scenario including also the result from LSND. Both studies will include all
further relevant exclusion limits from experiments with negative results and we
will see that CP-violation effects can in both cases be quite sizable.

In the following discussion of the three neutrino scenario we will
call $\nu_1$ and $\nu_2$ the mass eigenstates which correspond to 
the smallest $|\Delta m^2|$. The ATM and SUN results imply a hierarchy 
between the relevant squared mass differences\footnote{This
hierarchy of frequencies can correspond to a situation in which
one mass is much larger than the other two, $m_3\gg m_1,m_2$ and in
particular, but not necessarily, $m_3\gg m_2\gg m_1$, to a
situation in which two masses are degenerate, $m_1\simeq m_2\neq
m_3$ or to a completely degenerate situation $m_1\cong m_2\simeq m_3$.}
which leads to 
$|\Delta m^2_{12}|\ll|\Delta m^2_{23}|\simeq|\Delta m^2_{13}|$.
In this limit we can write for eq.~(\ref{3n})
\globallabel{3nh}
\begin{align}
\PCPC(\nu_{e_i}\rightarrow\nu_{e_j}) &\simeq \delta_{ij}
-4|U_{i3}|^2(\delta_{ij}-|U_{j3}|^2) \sin^2\Delta_{23}
-4\re(U^\pdagger_{j1}U^\dagger_{1i}U^\pdagger_{i2}U^\dagger_{2j})
\sin^2\Delta_{12} ~,\mytag \\
\PCPV(\nu_{e_i}\rightarrow\nu_{e_j}) &\simeq 8\,\sigma_{ij} 
J \sin^2\Delta_{23} \sin\Delta_{12}~.
\mytag 
\end{align}
The implications of solar and atmospheric neutrino experiments on the
parameter space are easy to recover due to the constraints on the
matrix element $U_{e3}$ between the electron neutrino and the third
mass eigenstate. First of all a large $|U_{e3}|^2$ (i.e. close to unity) 
is excluded because, due to unitarity, it would prevent solar neutrinos
from oscillating. Moreover a three neutrino fit of the atmospheric
neutrino data~\cite{barger:98a} gives $|U_{e3}|^2<0.08$ at 95\% CL. Finally
if $|\Delta m^2_{23}|\gtrsim 2\cdot 10^{-3}\eV^2$ the results of the
Chooz experiment~\cite{CHOOZ97}  give $|U_{e3}|^2<0.05$ at 90\%
CL. These constraints are strong enough to decouple the solar and
atmospheric neutrino analysis. In fact in the limit of small 
$|U_{e3}|^2$, namely small $\theta_{13}$ in the 
parameterization of eq.~(\ref{Ut1}), it is easy to
recover  from eqs.~(\ref{3nh}) that the oscillation
probability for solar neutrinos is\footnote{Eq.~(\ref{DecSol})
holds in vacuum, but the argument is essentially unchanged 
in presence of matter effects.}
\begin{equation}
\label{DecSol}
P(\nu_e\rightarrow\nu_e)\simeq 1 -4|U_{e1}U_{e2}|^2 \sin^2\Delta_{12}\simeq 1
-\sin^2 2\theta_{12} \sin^2\Delta_{12}~,
\end{equation}
so that the solar neutrino plots have to be read in the $\sin^2
2\theta_{12}$--$|\Delta m^2_{12}|$ plane. On the other hand the
probabilities involved in atmospheric neutrino experiments are
$P(\nu_e\rightarrow\nu_e)\simeq 1$, $P(\nu_e\rightarrow\nu_\mu)\simeq
P(\nu_\mu\rightarrow\nu_e)\simeq 0$~, 
\begin{equation}
\label{DecAtm}
P(\nu_\mu\rightarrow\nu_\mu)\simeq 1 -4|U_{\mu 3}|(1-|U_{\mu 3}|)^2
\sin^2\Delta_{23}\simeq 1 -\sin^2 2\theta_{23} \sin^2\Delta_{23}~,
\end{equation}
so that the atmospheric neutrino plots have to be read in the $\sin^2
2\theta_{23}$--$|\Delta m^2_{23}|$ plane.
Therefore~\cite{bahcall:98a,SK:98a} from the experimental data follows
\globallabel{Dm2}
\begin{align}
|\Delta m^2_{23}| &= \Delta m^2_{\text{ATM}} =
(0.3\mbox{--}8)\cdot10^{-3}\eV^2 \quad \text{atmospheric (ATM)}~,\mytag \\
|\Delta m^2_{12}| &= 
\Delta m^2_{\text{SUN}} =
\left\{
\begin{array}{l}
(0.2\mbox{--}2.0)\cdot10^{-4}\eV^2 \quad \text{large angle MSW (LMA)} \\
(0.4\mbox{--}1.0)\cdot10^{-5}\eV^2 \quad \text{small angle MSW (SMA)} \\
(0.5\mbox{--}2.0)\cdot10^{-10}\eV^2 \quad \text{vacuum}
\end{array}
\right.,
\mytag
\end{align}
as well as
$\sin^2 2\theta_{23}\gtrsim 0.8$ and $\sin^2 2\theta_{12}\gtrsim 0.7$
for the large angle MSW solution, all of them at 99\% CL. For given 
$\Delta m^2_{23}$ we can see from eq.~(\ref{3nh}\sep{b}) that $\PCPV$ 
becomes maximal if $\Delta m^2_{12}$ is in the MSW range for maximal $J$, 
i.e. maximal mixing angle. This is the LMA case and we can see now why it 
is especially interesting from the point of CP-violating effects in the
case of three neutrinos. The point is that this solution is neither 
affected by the small angle suppression of the SMA solution nor by the 
small $\Delta m^2_{\text{SUN}}$ suppression of the vacuum solution. 

The CP-violating asymmetry $a^{\CPC}_{ij}$, which depends on the 
parameters $|\Delta m^2_{12}|$, $|\Delta m^2_{23}|$, $\theta_{12}$, 
$\theta_{23}$, $\theta_{13}$ and $\delta$ in different ways, can now be 
analyzed. The largest impact on $a^{\CPC}_{ij}$ comes from $\theta_{13}$ 
and $|\Delta m^2_{12}|$ since $\theta_{13}$ can have any value between 
zero and the limit discussed above and $|\Delta m^2_{12}|$ is the most 
important suppression factor. $|\Delta m^2_{23}|$ is also important 
since it controls the relative importance of the $\sin^2\Delta_{12}$ 
terms compared to the $\sin^2\Delta_{23}$ 
terms in oscillation probabilities. On the other hand $\theta_{12}$ 
and $\theta_{23}$ do not affect $a^{\CPC}_{ij}$ very much since 
they are rather strongly constrained. Finally the $\delta$-dependence 
of $a^{\CPC}_{ij}$ is of course crucial. $a^{\CPC}_{ij}$ is however 
in leading approximation only linear in $\sin\delta$ and it makes 
therefore sense\footnote{The quantity $|a^\CPC_{\nu_\mu\nu_e}/\sin\delta|$ 
is independent of $\delta$ only in the limit in which the following 
eq.~(\ref{aem}) holds. The lower parts of the contour plots of 
$|a^\CPC_{\nu_\mu\nu_e}/\sin\delta|$ in fig.~\ref{fig:asy3me} 
have in fact a dependence on $\delta$, whereas the upper parts are 
almost $\delta$-independent. Fig.~\ref{fig:asy3me} assumes
$|\sin\delta|=1$. For smaller values of $|\sin\delta|$ the lower parts
of the plots in fig.~\ref{fig:asy3me} correspond to larger or smaller
values of $|a^\CPC_{\nu_\mu \nu_e}/\sin\delta|$ according to the sign
of $\cos\delta$ but at the same time $|a^\CPC_{\nu_\mu \nu_e}|$ gets
smaller. When $|\sin\delta|=1$ the plots do not either depend on the
angle $\theta_{12}$ corresponding to $\sin^2 2\theta_{12}=1$ 
or on the angle $\theta_{23}$ corresponding to 
$\sin^2 2\theta_{23}=1$.\label{foo:del}} 
to study $a^{\CPC}_{ij}/\sin\delta$ instead of $a^{\CPC}_{ij}$.
We will therefore discuss for the rest of this section in a 
quantitative and qualitative way the dependence of 
$|a^{\CPC}_{ij}/\sin\delta|$ on $\theta_{13}$, $|\Delta m^2_{12}|$ and
$|\Delta m^2_{23}|$ in the channels
$\nu_\mu\rightarrow\nu_e$/$\bar{\nu}_\mu\rightarrow\bar{\nu}_e$,
$\nu_\mu\rightarrow\nu_\tau$/$\bar{\nu}_\mu\rightarrow\bar{\nu}_\tau$,
of long-baseline experiments. These are in the case of three neutrinos 
the only experiments where CP-violating effects have a chance of being 
sizable. Short-baseline experiments can not be affected by CP-violating 
effects since the largest squared mass difference is of order $10^{-3}\eV^2$. 
Atmospheric neutrino experiments are in the case of three neutrinos (and we
will see also in the four neutrino case) also not much affected by
CP-violation, as we will explicitely see in a moment. The smallness of 
matter effects allows us, as mentioned before, to analyze the asymmetries 
first analytically in vacuum. A numerical determination of the matter 
effect corrections will follow at the end of this section.

\subsection{Long-baseline $\nu_\mu\rightarrow\nu_e$}

\noindent 
Let us consider the $\nu_\mu\rightarrow\nu_e$ channel and study
first the qualitative features of $a^\CPC_{ij}$ for the large
mixing angle solution. For $\sin^2 2\theta_{12}=\sin^2 2\theta_{23}=1$ 
one obtains
\globallabel{asy3me}
\begin{gather}
\PCPC(\nu_\mu\rightarrow\nu_e) \simeq 
\frac{\sin^2 2\theta_{13} \vev{\sin^2\Delta_{23}}}{2}~, \mytag\\
|\PCPV(\nu_\mu\rightarrow\nu_e)| = |\cos\theta_{13} \sin 2\theta_{13}
\sin\delta \vev{\sin^2\Delta_{23} \sin\Delta_{12}}| ~.\mytag 
\end{gather}
Even though a long-baseline experiment is not sensitive to the 
suppressed $|\Delta m^2_{12}|$ terms through the CP-conserving part of 
the probability (which is quadratic in $\sin\Delta_{12}$), it can still 
be sensitive to $|\Delta m^2_{12}|$ through the CP-violating term for 
two reasons: First $\PCPV$ is only linearly suppressed by 
$\sin\Delta_{12}$. Second $\PCPC$ 
is suppressed by $\sin^2 2\theta_{13}$ (which can not be large), whereas
$\PCPV$ contains only $\sin 2\theta_{13}$, so that experiments able
to detect a CP-conserving probability which is twice suppressed by
$\sin\theta_{13}$ can see larger asymmetries. From eqs.~(\ref{asy3me}) 
follows 
\begin{equation}
\left|a^\CPC_{\nu_\mu \nu_e}\right| \simeq \left|\frac{2}{\sin 2\theta_{13}}
\frac{\vev{\sin^2\Delta_{23} \sin\Delta_{12}}}
{\vev{\sin^2\Delta_{23}}}\sin\delta \right|~,
\label{aem}
\end{equation}
that shows how the $\vev{\sin\Delta_{12}}$ suppression is balanced by the 
$1/\sin 2\theta_{13}$ enhancement of $a^\CPC_{\nu_\mu \nu_e}$.
The weight function $f_+$ (i.e. experimental details) enters in the 
asymmetry $a^\CPC$ to a good approximation only via $\vev{L/E}$ in
$\vev{\sin\Delta_{12}}\simeq \Delta m^2_{12}/4 \vev{L/E}$. 

\begin{figure}
\begin{center}
\epsfig{file=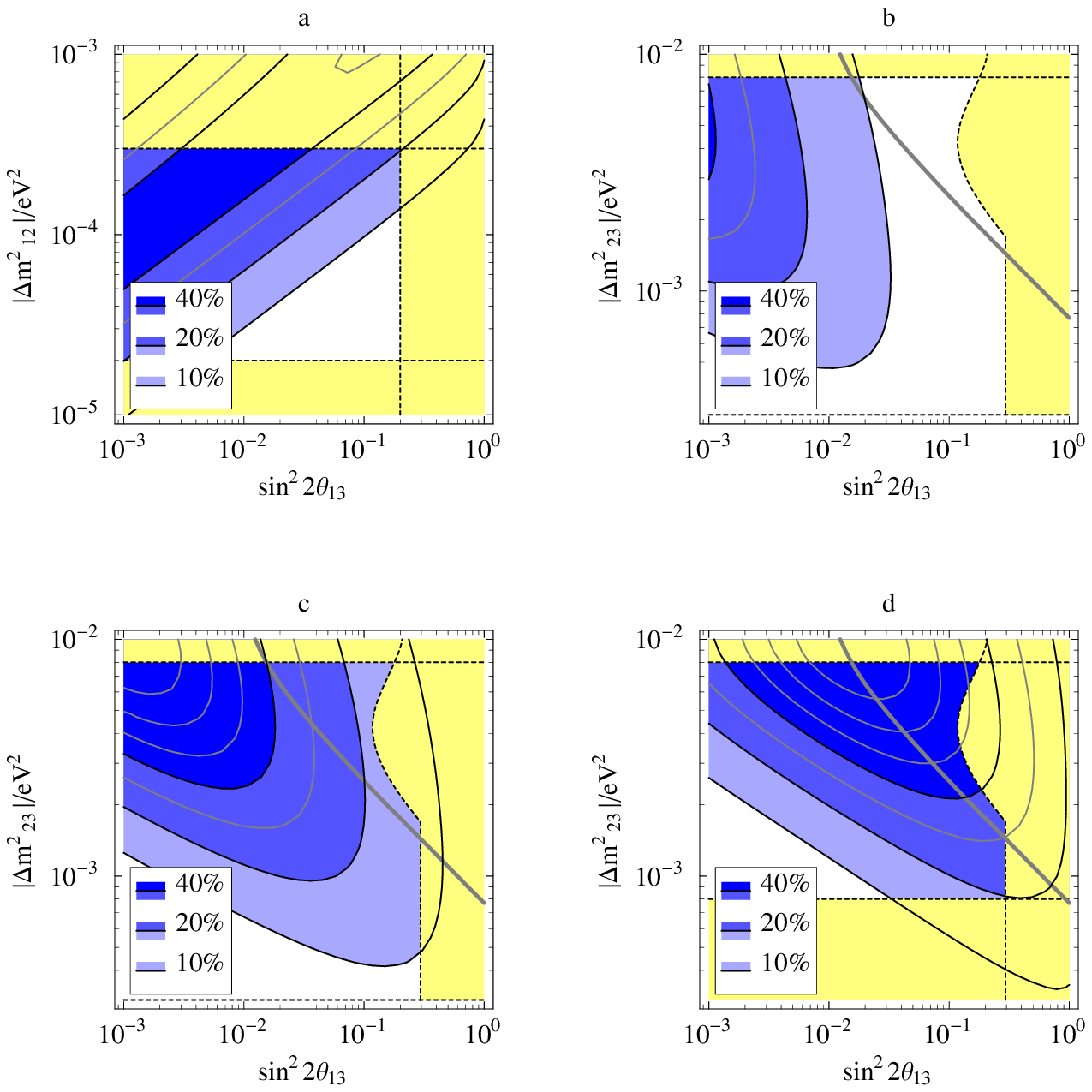,width=0.9\textwidth}
\end{center}
\mycaption{Contour lines for
$|a^\CPC_{\nu_\mu \nu_e}/\sin\delta|$ in the $\sin^2 2\theta_{13}$--$|\Delta
m^2_{12}|$  plane for $|\Delta m^2_{23}| = 2\cdot~10^{-3}\eV^2$ \textrm{(a)}
and in the $\sin^2 2\theta_{13}$--$|\Delta m^2_{23}|$ plane for $|\Delta
m^2_{12}| = 0.5\cdot 10^{-4}\eV^2, 2\cdot 10^{-4}\eV^2, 8\cdot~10^{-4}\eV^2$ 
\textrm{(b,c,d)} for $\sin^2
2\theta_{12} = \sin^2 2\theta_{23} = 1$ and $\vev{L/E}\simeq 100
\text{\textrm{m}}/\MeV$ (see text). The horizontal shadowed regions limit
the range of $|\Delta m^2|$ according to eqs.~(\ref{Dm2}) (and $|\Delta
m^2_{23}|>|\Delta m^2_{12}|$ in (d)) whereas the shadowed regions on
the right correspond to the combined Chooz and atmospheric
neutrino limits. The thick solid line represents the
MINOS sensitivity.}
\label{fig:asy3me}
\end{figure}

Fig.~\ref{fig:asy3me} shows the contour lines for $|a^\CPC_{\nu_\mu
  \nu_e}/\sin\delta|$ in the $|\Delta m^2_{12}|$--$\sin^2 2\theta_{13}$ plane
for fixed $|\Delta m^2_{23}|$ (fig.~\ref{fig:asy3me}a) and in the $|\Delta
m^2_{23}|$--$\sin^2 2\theta_{13}$ plane for fixed $|\Delta m^2_{12}|$
(fig.~\ref{fig:asy3me}b,c,d) using the exact formulas for the oscillation
probability in eqs.~(\ref{3n}) and setting $\sin^2 2\theta_{12} = \sin^2
2\theta_{23} = 1$, which as mentioned does not affect the generality of our
results. We assumed an experiment with a baseline of $L=730\text{km}$ and a
energy distribution around $E=6\GeV$ with $\sigma_E\sim3\GeV$, giving
$\vev{L/E}\sim 100\text{m}/\MeV$. This looks like the MINOS \cite{minos}
setup\footnote{The MINOS experiment can run in different
  configurations. The energy distribution used here refers to the so called
  ``PH2(medium)'' initial neutrino flux, since it gives a sensitivity
  comparable to the PH2(high) flux but a better value of $\vev{L/E}$.}, but 
the results depend to a good approximation only on $\vev{L/E}$. For other
experiments with different values of $\vev{L/E}$ one can therefore rescale the
asymmetries of fig.~\ref{fig:asy3me} with a factor
$\vev{L/E}/(100\text{m}/\MeV)$.  The horizontal shadowed regions limit the
range of $|\Delta m^2|$ according to the values given in eq.~(\ref{Dm2}) (and
$|\Delta m^2_{23}|>|\Delta m^2_{12}|$ in fig.~\ref{fig:asy3me}d). The MINOS
sensitivity taken from ref.~\cite{minos} (thick solid line) and the region excluded by Chooz and the atmospheric
neutrino fits (vertically shaded regions) are also displayed. The figures 
show that the parameter space which is accessible by the MINOS experiment 
is not that large. Nevertheless there could be maximally a 30--40\% effect.

The general structure of figs.~\ref{fig:asy3me} can be easily understood
with the help of the approximations yielding eq.~(\ref{aem}).
Fig.~\ref{fig:asy3me}a shows that for maximal CP-violation 
($|\sin\delta|=1$) the asymmetry can easily reach rather large values. 
Of course larger values of $|\Delta m^2_{12}|$ are preferred. As long 
as $\sin^2 2\theta_{13}$ is not too small (otherwise the $\sin^2\Delta_{23}$ 
terms in eq.~(\ref{3n}\sep{b}) become as small as the $\sin^2\Delta_{12}$ ones
and eqs.~(\ref{asy3me}) is no longer valid) also smaller values of $\sin^2
2\theta_{13}$ are preferred, as expected. Figs.~\ref{fig:asy3me}b,c,d 
show the dependence on $|\Delta m^2_{23}|$ in greater detail. 

As long as the approximations leading to eq.~(\ref{aem}) hold,
$|a^\CPC_{\nu_\mu \nu_e}/\sin\delta|\propto (\sin^2 2\theta_{13})^{-1/2}$ 
for fixed $|\Delta m^2_{12}|$ and the contours are vertical a part from the 
(small) effects of $\sin^2\Delta_{23}$ in
$\vev{\sin^2\Delta_{23} \sin\Delta_{12}}/\vev{\sin^2\Delta_{23}}$. On the 
other hand, for lower
values of $|\Delta m^2_{23}|$ the $\sin^2\Delta_{12}$ terms in the
CP-conserving part of the probability are not negligible anymore
(otherwise eq.~(\ref{aem}) would give $|a^\CPC_{\nu_\mu\nu_e}| >1$ for 
$\sin^2 2\theta_{13}$ low enough). The value of $|\Delta m^2_{23}|$ at 
which the $\sin^2\Delta_{12}$ terms fold the contour-lines is higher when the
$\sin^2\Delta_{23}$ term in $\PCPC$ is smaller, namely when $\sin^2
2\theta_{13}$ is smaller, on the left-side of the plot.

In fig.~\ref{fig:asy3me}b a ``CP-disfavouring'' value of 
$|\Delta m^2_{12}|=0.5\cdot 10^{-4}$ has been chosen, while in 
fig.~\ref{fig:asy3me}c the ``CP-optimistic'' value 
$|\Delta m^2_{12}|=2\cdot 10^{-4}$ has been used. 
In fig.~\ref{fig:asy3me}d a value of $|\Delta m^2_{12}|$
possible only in some non-standard solar analysis (see below) and
therefore lying in the shadowed region in fig.~\ref{fig:asy3me}a has
been considered: $|\Delta m^2_{12}|=8\cdot 10^{-4}$.  We see from
figs.~\ref{fig:asy3me}b,c that in case of maximal CP-violation the
effects in the allowed region covered by the MINOS sensitivity range
from 2\% to 10\% ($|\Delta m^2_{12}| = 0.5\cdot 10^{-4}\eV^2$) and
from 10\% to 40\% ($|\Delta m^2_{12}| = 2\cdot 10^{-4}\eV^2$),
according to the value of $|\Delta m^2_{23}|$ and the amplitude of the
$\nu_\mu\rightarrow\nu_e$ oscillation.

Even in the framework of a standard analysis of solar data, these effects 
can therefore be large enough to spoil the analysis of a possible 
signal measured by a long-baseline experiment, if this analysis 
does not take into account CP-violation. In figs~\ref{fig:asy3me} 
this can be seen explicitely from the shown sensitivity goals of MINOS.
Regarding the possibility of measuring such an asymmetry, it should 
be noticed that the asymmetry is larger when the amplitude is smaller,
so that an enhancement of the CP-violation effect is accompanied by a
suppression of the statistics. Thus there is no advantage from a statistical
point of view. Large asymmetries should make it however easier to distinguish
intrinsic from experimental asymmetries. 

In some versions of solar neutrino data analysis $|\Delta m^2_{12}|$ 
can lie in the lower part of the atmospheric range. This can happen, 
for example, if one assumes an unknown large systematic error in the 
Chlorine experiment~\cite{BHSSW}. This is interesting, since the 
exclusion of the Chlorine data from the analysis makes the remaining 
solar data also consistent with an energy independent reduction of 
the solar flux as it happens to be above the MSW range and
for almost maximal oscillation amplitude. As a consequence, for 
$\sin^2 2\theta_{13} \lesssim 0.25$, as given by the Super-Kamiokande 
atmospheric analysis, the large angle solution becomes a vertical 
strip in the $|\Delta m^2_{12}|$-$\sin^2 2\theta_{12}$ plane close 
to the $\sin^2 2\theta_{12} =1$ axis and extending to the Chooz 
limit\footnote{When $|\Delta m^2_{12}|$ approaches the Super-Kamiokande 
atmospheric range, the $|\Delta m^2_{12}|$ effects become important 
in the atmospheric analysis so that the constraint
$\sin^2 2\theta_{13} \lesssim 0.25$, as well as $\sin^2
2\theta_{23}\simeq 1$ is not reliable anymore.}
$|\Delta m^2_{12}| < 10^{-3}\eV^2$. This explains why we considered 
in fig.~\ref{fig:asy3me}d  the possibility $|\Delta
m^2_{12}|=8\cdot 10^{-4}\eV^2$, where the asymmetry is always large
for maximal phases\footnote{A non-standard solar analysis leaves
open the possibility that $|\Delta m^2_{12}|$ falls in the atmospheric
range while remaining under the Chooz constraint. One can then wonder
whether \textit{i)} $|\Delta m^2_{12}|$ could be responsible for both
the solar (Chlorine excluded) and atmospheric evidences with $|\Delta
m^2_{23}|$ above the atmospheric range and whether \textit{ii)}
$|\Delta m^2_{23}|$ could explain the LSND signal.  The answers are
that, independently of \textit{ii)}, \textit{i)} is strongly
disfavoured and in any case \textit{ii)} is not possible. See appendix A}.

It is interesting to observe that the CP-violating part in the
$\nu_\mu\rightarrow\nu_e$ amplitudes at long-baseline experiments can provide
information on parameters typically accessible to solar neutrino experiments 
(at least in the present 3-neutrino scenario), while the
CP-conserving part of the amplitudes are almost insensitive to the same
parameters. The detection of a CP-violation effect would for example select 
the large angle MSW solution out of the three possible solutions of the solar
neutrino problem. The values of $|a^\CPC_{\nu_\mu \nu_e}|$ and $\sin^2
2\theta_{13}$, from the CP-conserving part, would select a range for $|\Delta
m^2_{12}|$.  This can be understood with the help of fig.~\ref{fig:asy3me}a.
The measurement of $\sin^2 2\theta_{13}$ would select a vertical line in that
plot. On the other hand, since that figure is ``maximal'', namely plotted for
$|\sin\delta| =1$, values of $|a^\CPC_{\nu_\mu \nu_e}|$ in this figure lower 
than the measured one would be ruled out, allowing thus only a limited range 
for $|\Delta m^2_{12}|$. Since lower values of $|\Delta m^2_{12}|$ would give
a small CP-violation effect, such a measurement would select the upper part 
of the $|\Delta m^2_{12}|$ range provided by the solar analysis and larger
asymmetries would correspond to stronger lower limits on $|\Delta m^2_{12}|$.

The parameters of our typical long baseline experiment are not 
too far away from the MSW resonance region. We must therefore 
include matter effects in our discussion as outlined in 
subsection~\ref{subsec:asymmetries} and we will now demonstrate 
that the corrections to the results obtained so far are moderate.
Matter effects lead as usual to an inherent sensitivity of the
charged current interactions of neutrinos to the electron flavour 
and diagonalization of the Hamiltonian leads to new mass eigenstates 
and shifted masses for the propagation in matter. The transition 
from vacuum to matter can altogether be phrased as two parameter mappings 
\begin{equation}
(\theta_{12},\theta_{13},\theta_{23},\delta,\Delta m^2_{12},\Delta m^2_{23})
\rightarrow (\theta_{12}',\theta_{13}',\theta_{23}',\delta',(\Delta
m^2_{12})',(\Delta m^2_{23})') ~,
\end{equation}
one for neutrinos and one for antineutrinos. In order to determine the 
corrections due to matter effects we studied for $f = \bar f$ numerically 
the asymmetry $a^{\mathrm{CP}}_{\mu e}$ for the following cases:
a) $a^{\mathrm{CP}}_{\mu e}\equiv a^{\mathrm{tot}}_{\mu e}$ in vacuum 
   for maximal CP-phase,
b) $a^{\mathrm{tot}}_{\mu e}$ in matter with maximal phase, 
c) $a^{\mathrm{exp}}_{\mu e}$ in matter and 
d) $a^{\mathrm{CP}}_{\mu e}$ in matter with maximal phase.
\begin{figure}
\begin{center}
\epsfig{file=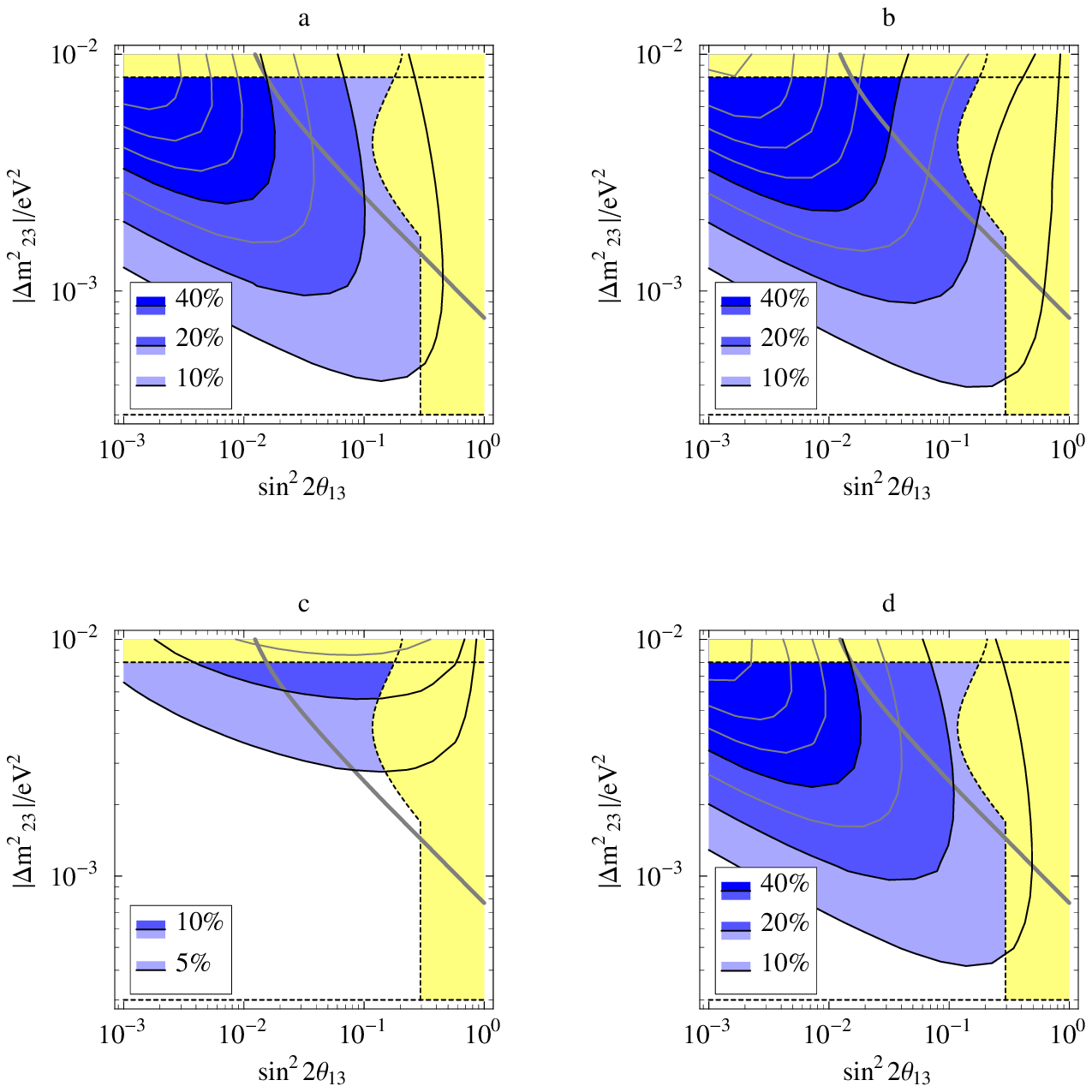,width=0.9\textwidth}
\end{center}
\mycaption{
CP-asymmetries for the four cases discussed in the text in analogy to fig.~1:
(a) $a^{\mathrm{CP}}_{\mu e}\equiv a^{\mathrm{tot}}_{\mu e}$ 
   in vacuum for maximal CP-phase, 
(b) $a^{\mathrm{tot}}_{\mu e}$ in matter with maximal phase, 
(c) $a^{\mathrm{exp}}_{\mu e}$ in matter and 
(d) $a^{\mathrm{CP}}_{\mu e}$ in matter with maximal phase.} 
\label{fig:asy3memsw}
\end{figure}
These asymmetries are plotted in figs.~\ref{fig:asy3memsw}a--d
for the $\nu_\mu\rightarrow\nu_e$ channel, where we use the scenario 
previously assumed for fig.~\ref{fig:asy3me}c, namely a LBL experiment 
with $\vev{L/E} \simeq 100\mathrm{m}/\MeV$, $\delta m^2_{12}=2\cdot
10^{-4}\eV^2$ and $\sin^2 2\theta_{12}=\sin^2 2\theta_{23}=1$.
Figure~\ref{fig:asy3memsw}b shows the CP--asymmetry measured in 
the experiment, plot (c) the matter induced asymmetry and (d) the 
experimental asymmetry with the matter induced asymmetry subtracted. 
For a few parameter points these asymmetries were already calculated 
in \cite{gavela98} and our figures agree perfectly in those points.
Note that matter effects depend both on the sign of $\Delta m^2$ and the 
sign of the CP-phase $\delta$. In fig.~\ref{fig:asy3memsw} we used a positive
$\delta$ and a positive $\Delta m^2$, in which case matter and intrinsic effects
go into the same direction.

Altogether we can see that matter effects require a generalization 
of the asymmetry in order to isolate genuine CP--violating effects 
from matter induced effects. Figures~\ref{fig:asy3memsw} show that 
the corrections are altogether moderate. Ideally one would like to 
study case d) which is due to intrinsic CP--violating effects, but 
would require to determine somehow $a^{\mathrm{exp}}_{\mu e}$ in 
matter. With good knowledge of the masses and mixing angles this may 
for example be possible by calculating this matter-induced asymmetry 
theoretically. To do so one has to use fig.~\ref{fig:asy3memsw}b 
and subtract (with the correct sign) the theoretically
calculated fig.~\ref{fig:asy3memsw}c, which leads to larger 
systematic uncertainties. For fig.~\ref{fig:asy3memsw}b the region 
with large CP-effects shifts to larger $\sin^2 2\theta_{13}$ and 
smaller $\Delta m^2_{23}$ and the maximal CP-asymmetry expected in 
the sensitivity range of MINOS increases from  40\% to 50\%.
This shows that the seperation is in principle possible, but 
it is clearly a difficult task which requires sufficient experimental 
information. Another promising method to seperate the matter-induced 
and intrinsic CP-asymmetries by using envelope patterns of the oscillation 
is discussed by Arafune, Koike and Sato \cite{others}.

\subsection{Long-baseline $\nu_\mu\rightarrow\nu_\tau$}

\begin{figure}
\begin{center}
\epsfig{file=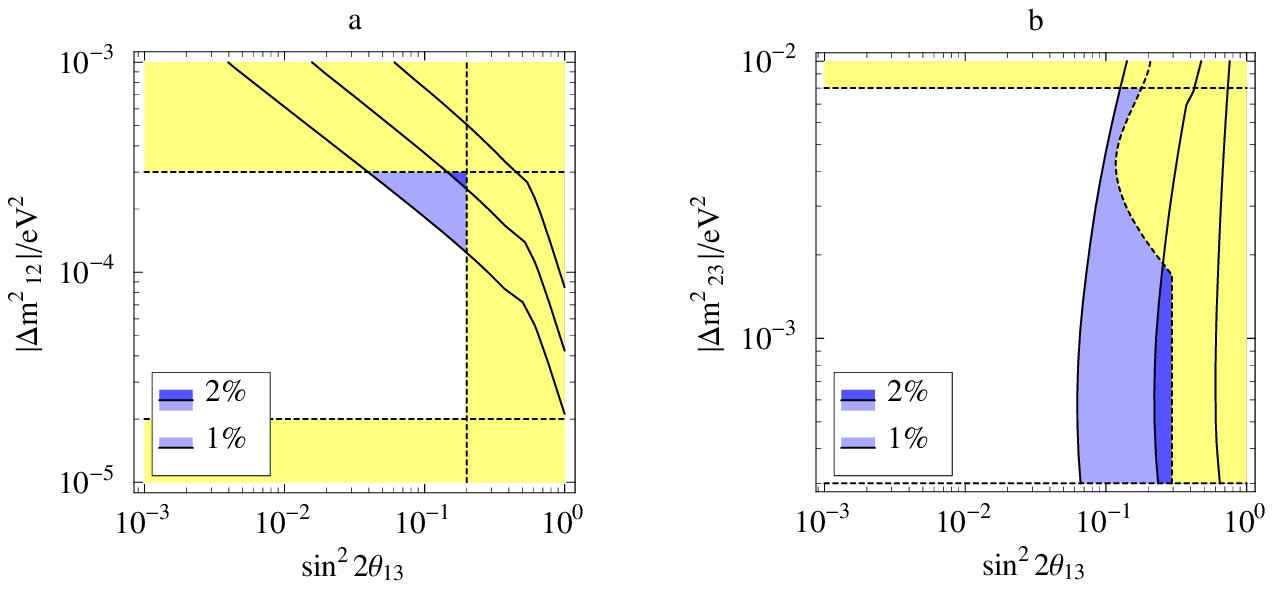,width=0.9\textwidth}
\end{center}
\mycaption{Contour lines for
$|a^\CPC_{\nu_\mu \nu_\tau}/\sin\delta|$ in the $|\Delta
m^2_{12}|$--$\sin^2 2\theta_{13}$ plane for $|\Delta
m^2_{23}| = 2\cdot~10^{-3}\eV^2$ \textrm{(a)} and in the $|\Delta
m^2_{23}|$--$\sin^2 2\theta_{13}$ plane for $|\Delta
m^2_{12}| = 2\cdot 10^{-4}\eV^2$ \textrm{(b)} for $\sin^2
2\theta_{12} = \sin^2 2\theta_{23} = 1$ and $\vev{L/E}\simeq 290
\text{\textrm{m}}/\MeV$ (see text). The horizontal shadowed regions limit
the range of $|\Delta m^2|$ according to eqs.~(\ref{Dm2}) (and $|\Delta
m^2_{23}|>|\Delta m^2_{12}|$ in (d)) whereas the shadowed regions on
the right correspond to the combined Chooz and atmospheric
neutrino limit.}
\label{fig:asy3mu}
\end{figure}

In the $\nu_\mu\rightarrow\nu_\tau$ channel the CP-violating
probability is the same as in the $\nu_\mu\rightarrow\nu_e$ channel (a
part from the sign) whereas the CP-conserving probability is not
suppressed by $\sin^2 2\theta_{13}$ anymore and is therefore
larger. Therefore the asymmetry is smaller in this channel. On one
hand the enhancement of the CP-conserving probability gives better
statistics and hence in principle the possibility to appreciate a smaller
asymmetry. This enhancement would on the other hand also
enhance the ``experimental'' contribution $a^{\text{exp}}$ to the
asymmetry, making it very hard to identify the CP-violating
contribution $a^\CPC$. The $\nu_\mu\rightarrow\nu_\tau$ channel 
is therefore essentially unsuitable for the detection of CP-violating 
effects, and one can neglected them in the analysis of a 
possible signal.
This statement is confirmed by fig.~\ref{fig:asy3mu}, where the 
contour lines for $|a^\CPC_{\nu_\mu \nu_\tau}/\sin\delta|$ 
are plotted in the $|\Delta m^2_{12}|$--$\sin^2 2\theta_{13}$ plane 
for $|\Delta m^2_{23}| = 2\cdot 10^{-3}\eV^2$ (a) and in the 
$|\Delta m^2_{23}|$--$\sin^2 2\theta_{13}$ plane for the 
optimistic case $|\Delta m^2_{12}|=2\cdot 10^{-4}\eV^2$ (b). 
The figures show that the asymmetries are always smaller than 5\%.
Matter effects may then (via its asymmetric influnece on the mass 
eigenvalues) even dominate the asymmetry. Therefore we 
do not consider the $\mu$--$\tau$ channel any further.

\section{Scenarios with four neutrinos\label{sec:4nu}}

In order to accommodate the LSND signal in addition to the solar and 
atmospheric results we will study in this section four neutrinos
scenarios. We expect that matter effects in long baseline experiments 
are moderate corrections like in the case of three neutrinos. The 
discussion of matter effects will therefore be covered elsewhere.
Let us first consider the possible mass ordering schemes 
resulting from the hierarchical values of the squared mass differences 
$\Delta m^2_{\text{SUN}} \ll  \Delta m^2_{\text{ATM}} \ll 
\Delta m^2_{\text{LSND}}$.
As in the case of three neutrinos we associate the smallest squared 
mass difference with the first two mass eigenstates, i.e.
$|\Delta m^2_{12}|=\Delta m^2_{\text{SUN}}$. Then there are only two 
possibilities for the ordering of the other mass differences relative 
to these first two states:

\begin{itemize}
\item[A)] The intermediate squared mass difference occurs between the
3rd
and 4th eigenstates, i.e. $|\Delta m^2_{34}|=\Delta m^2_{\text{ATM}}$
and the larger LSND value defines the splitting between the 1st/2nd and
3rd/4th mass eigenstate. Altogether this implies
$|\Delta m^2_{12}| \equiv \Delta m^2_{\text{SUN}} \ll |\Delta m^2_{34}|
\equiv \Delta m^2_{\text{ATM}} \ll |\Delta m^2_{14}| \simeq
|\Delta m^2_{24}| \simeq |\Delta m^2_{13}|
\simeq |\Delta m^2_{23}| \equiv \Delta m^2_{\text{LSND}}$
\item[B)] The intermediate squared mass difference occurs between one
of the 1st and 2nd eigenstate and one eigenstate
out of the 3rd and 4th. Conventionally they are the 2st and 3rd
mass eigenstate, i.e. $|\Delta m^2_{23}|=\Delta m^2_{\text{ATM}}$
and the larger LSND value defines the splitting between the
4th mass eigenstate and the others, i.e. $|\Delta m^2_{34}|=\Delta
m^2_{\text{LSND}}$. We have thus in this case
$|\Delta m^2_{12}| \equiv \Delta m^2_{\text{SUN}} \ll |\Delta m^2_{13}|
\simeq |\Delta m^2_{23}| \equiv \Delta m^2_{\text{ATM}} \ll
|\Delta m^2_{14}| \simeq |\Delta m^2_{24}| \simeq |\Delta m^2_{34}|
\equiv \Delta m^2_{\text{LSND}}$
\end{itemize}

Note that only $\Delta m^2$ enters in neutrino oscillation experiments 
and that this leaves some freedom in the ordering and absolute values 
of masses. Scheme B turns out to be in disagreement with experimental
data~\cite{bilenkii:98b}. We will therefore only consider scheme A in 
the following.
This can be understood in a simplified picture where only two neutrino 
mass eigenstates (i.e. their $\Delta m^2$) participate in each oscillation 
experiment together with the information about the involved flavour 
transitions of each experiment. If one starts in scenario B with 
$\Delta m^2_{\text{SUN}}$ as the smallest quadratic mass splitting which 
involves $\nu_e$ and assumes that $\Delta m^2_{\text{ATM}}$ (which 
fixes $\nu_\mu$) comes next, then the third and largest 
$\Delta m^2_{\text{LSND}}$ could not be any longer an oscillation 
between $\nu_\mu$ and $\nu_e$, which is a contradiction with the LSND 
experiment. 

Scenarios with four neutrinos involve in general a larger number of 
parameters in the mixing matrix with considerably more complexity in 
the parameter restrictions. The observed mass hierarchies allow however 
in experiments sensitive to $\Delta m^2\gtrsim 10^{-3}\eV^2$ the approximation 
$|\Delta m^2_{12}|\simeq 0$, unless $|\Delta m^2_{12}|$ is at the upper 
border or beyond its standard range, which will not be 
considered here. This approximation simplifies the general task 
considerably and reduces the number of involved parameters. Unlike 
the three neutrino case, in which $|\Delta m^2_{12}|$ could quite 
safely be neglected in the CP-conserving part of the probabilities while
it was crucial in the CP-violating ones, $|\Delta m^2_{12}|$ can be
safely neglected here completely. The oscillation probabilities are in 
the limit $|\Delta m^2_{12}|=0$ given by
\globallabel{12}
\begin{align}
\begin{split}
\PCPC(\nu_{e_i}\rightarrow\nu_{e_j}) &= \delta_{ij}(1 -4|U_{i4}|^2
\sin^2\Delta_{24} -4|U_{i3}|^2 \sin^2\Delta_{23}) \hfill \\
&-4\re J^{e_j e_i}_{34} \sin^2\Delta_{34} 
+4(\re J^{e_j e_i}_{34} +\re J^{e_j e_i}_{44}) \sin^2\Delta_{24} \hfill \\
&+4(\re J^{e_j e_i}_{33} +\re J^{e_j e_i}_{34}) \sin^2\Delta_{23}~,
\end{split} \mytag \\
\PCPV (\nu_{e_i}\rightarrow\nu_{e_j}) &= 8 \im J^{e_j e_i}_{34}
(\sin^2\Delta_{23}\sin\Delta_{34}\cos\Delta_{34}
+\sin^2\Delta_{34}\sin\Delta_{23}\cos\Delta_{23})~, \mytag
\end{align}
where the second and third line of~(\ref{12}\sep{a}) are of special interest 
for our purposes.  

Eq.~(\ref{12}b) shows that it is possible to generate the CP-violating 
part of the probabilities from $|\Delta m^2_{23}|$ and $|\Delta m^2_{34}|$.
CP-violation in long-baseline experiments is therefore no longer
suppressed by the small $\Delta m^2_{\text{SUN}}$ as in the three neutrino case
and we will see that it can therefore be large. With four neutrinos one 
can also wonder, whether CP-violation can be important in short-baseline 
experiments able to measure small transition probabilities. 
We will see that, although the effects are not large, a modest improvement of
$\vev{L/E}$ would be enough to see sizeable effects in the
$\nu_\mu\rightarrow\nu_\tau$ channel, if the CP-violation phase is large. This
is because the relative importance of CP-violation becomes larger for smaller
effects, unlike what happens in the $\nu_\mu\rightarrow\nu_e$ case, where the
relative importance of CP-violation is always small.

Concerning the possibility of CP-violation effects in atmospheric neutrino
experiments, we notice that in the four neutrino case they are even more
unlikely then in the three neutrino one.  One may wonder why the four neutrino
case does not contain the three neutrino scenario as a specific limit. This is
however the case since there is one additional large frequency and also further
constraints from experiments sensitive to that frequency. Especially important
is here the constraint on $\nu_e\rightarrow\nu_e$ 
from the Bugey experiment which guarantees in this scenario that
$\nu_\mu\leftrightarrow\nu_e$ oscillations (and therefore CP-violation, which
appears only there) do not play a role in atmospheric neutrino oscillations. As
a consequence CP-violation is negligible in atmospheric neutrino oscillations
and we will therefore consider in the following only long- and short-baseline
$\nu_\mu\rightarrow\nu_e$ and $\nu_\mu\rightarrow\nu_\tau$ experiments.

From eqs.~(\ref{12}) we can see that the oscillation probabilities between two
different flavour eigenstates $\nu_{e_i}$ and $\nu_{e_j}$ depend only on the
$2\times 2$ sub-sector of the mixing matrix involving the $i$th and $j$th
flavour eigenstates and the 3rd and 4th mass eigenstates. That sub-matrix is
described by 8 real parameters among which 3 are unphysical phases that can be
rotated away, one is a physical phase and 4 are mixing parameters. We choose
the 5 physical parameters as follows: Let $\nu'_{e_i} = U_{i3}\nu_3 +
U_{i4}\nu_4$, $\nu'_{e_j} = U_{j3}\nu_3 + U_{j4}\nu_4$ be the projections of
the flavour eigenstates $\nu_{e_i}$ and $\nu_{e_j}$ on the 3--4 mass
eigenspace. Then we define analogous to ref.~\cite{bilenkii:98a} the quantities
$c_i$ and $c_j$ as the squared lengths of these projections, i.e.
$c_i=|U_{i3}|^2+|U_{i4}|^2$, $c_j=|U_{j3}|^2+|U_{j4}|^2$. Furthermore
$\theta_i\in[0,\pi/2]$ is the orientation of $\nu'_{e_i}$ defined via
$(|U_{i3}|,|U_{i4}|)=\sqrt{c_i}(\cos\theta_i,\cos\theta_j)$ and
$\phi_{ij}\in[0,\pi/2]$ is the relative orientation of $\nu'_{e_i}$ and
$\nu'_{e_j}$, $\cos\phi_{ij}=|U^*_{i3}U_{j3}+U^*_{i4}U_{j4}|/\sqrt{c_i c_j}$.
Finally we define the relevant CP-violating phase as
$\delta_{ij}=\mathrm{arg}(U^*_{i3}U_{j3}U^*_{j4}U_{i4})= \mathrm{arg}(J^{e_j
  e_i}_{34})$.  The amplitudes of the oscillating terms in eqs.~(\ref{12}) can
be expressed by these parameters to be \globallabel{12par}
\begin{gather}
4\re J^{e_j e_i}_{34} = c_i c_j\sin 2\theta_i (\cos 2\phi_{ij}\sin 2\theta_i
+\sin 2\phi_{ij}\cos 2\theta_i\cos\delta_{ij})~, \mytag \\
4(\re J^{e_j e_i}_{34} +\re J^{e_j e_i}_{44}) = 4 c_i
c_j\sin\theta_i\cos\phi_{ij}
(\cos\phi_{ij}\sin\theta_i+\sin\phi_{ij}\cos\theta_i\cos\delta_{ij})~,
\mytag \\
4(\re J^{e_j e_i}_{33} +\re J^{e_j e_i}_{34}) = 4 c_i
c_j\cos\theta_i\cos\phi_{ij}
(\cos\phi_{ij}\cos\theta_i-\sin\phi_{ij}\sin\theta_i\cos\delta_{ij})~,
\mytag \\
4\im J^{e_j e_i}_{34} = c_i c_j \sin 2\theta_i\sin
2\phi_{ij}\sin\delta_{ij}~.
\mytag
\end{gather}
The introduced parameters $c_i$, $c_j$, $\theta_i$, $\phi_{ij}$, 
$\delta_{ij}$ can however not be chosen arbitrary since the $ij$/34 
sub-matrix is part of a unitary matrix. The parameters must fulfill
the following unitary constraints:
\globallabel{4uni}
\begin{gather}
0\leq c_i\leq 1 ~; \qquad 0\leq c_j\leq 1 ~,\mytag \\
c_i c_j\cos^2\phi_{ij} \leq (1-c_i)(1-c_j)~. \mytag
\end{gather}
These conditions can be easily understood noticing that the minor can 
be embedded in an unitary matrix if and only if the two $\mathbb{C}^2$ 
vectors $(U_{i3},U_{i4})$ and $(U_{j3},U_{j4})$ can be completed 
to a pair of orthonormal $\mathbb{C}^4$ vectors. Eq.~(\ref{4uni}\sep{a}) 
corresponds then to the normalization condition and 
eq.~(\ref{4uni}\sep{a}) to the orthogonality condition of the two 
$\mathbb{C}^4$ vectors.

\begin{figure}[htb]
\hspace*{0.225\textwidth}\epsfig{file=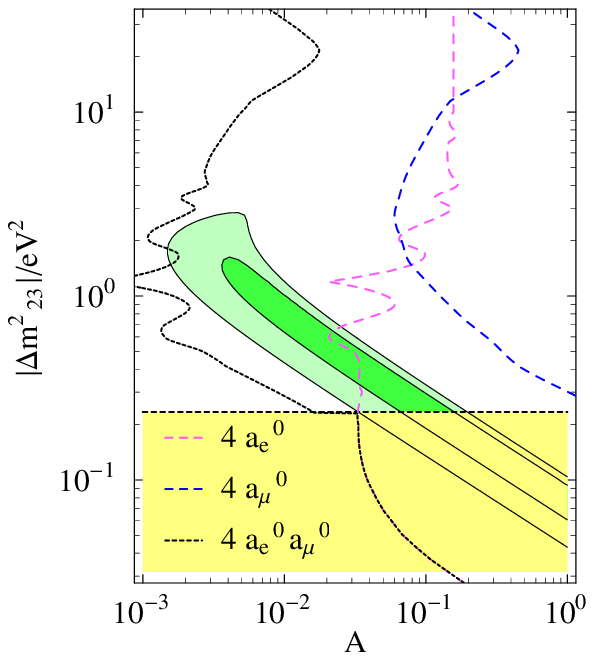,width=0.50\textwidth}
\mycaption{Constraints on the amplitude $A=4c_ec_\mu\cos^2\phi_{\mu e}$
of the LSND transition probability. The shown allowed regions in the 
LSND parameter space correspond to 90\% and 99\% C.L. The upper part 
of the plot is not shown because it is excluded by BNL E734.
The dashed lines represent $4a^0_e$ (lighter) and $4a^0_\mu$ (darker),
whereas the dotted line represents $4 a^0_e a^0_\mu$. The shadowed
region is excluded by the condition~(\ref{coA}).}
\label{fig:LSND}
\end{figure}

The parameters $c_i$, $c_j$, $\theta_i$, $\phi_{ij}$, $\delta_{ij}$ are not
only constrained by unitarity, but also by the results of the
$\nu_e\rightarrow\nu_e$ and $\nu_\mu\rightarrow\nu_\mu$ disappearance,
atmospheric neutrinos and LSND experiments. 
In order to make these constraints explicit, let us write the formulae
for the relevant processes in the approximation where $|\Delta m^2_{34}| \ll
|\Delta m^2_{23}| \simeq |\Delta m^2_{24}|$: 
\globallabel{4phen}
\begin{align}
P(\nu_e\rightarrow\nu_e) & \simeq 
1 -4 c_e (1-c_e) \sin^2\Delta_{23} -c_e^2\sin^2 2\theta_e \sin^2\Delta_{34}~,
\mytag \\ 
P(\nu_\mu\rightarrow\nu_\mu) & \simeq 
1 -4 c_\mu (1-c_\mu) \sin^2\Delta_{23} 
-c_\mu^2\sin^2 2\theta_\mu \sin^2\Delta_{34}~,
\mytag \\
\begin{split}
P(\nu_\mu\rightarrow\nu_e) & \simeq 4 c_e c_\mu \cos^2\phi_{\mu e}
\sin^2\Delta_{23} +2 c_e c_\mu 
\sin 2\theta_\mu\sin 2\phi_{\mu e}\sin\delta_{\mu e} 
\sin^2\Delta_{23}\sin\Delta_{34} \\
&\quad -c_e c_\mu \sin 2\theta_\mu (\cos 2\phi_{\mu e}\sin
2\theta_\mu +\sin 2\phi_{\mu e}\cos 2\theta_\mu\cos\delta_{\mu e})
\sin^2\Delta_{34}~.
\end{split}
\mytag
\end{align}
First we notice that only the $\sin^2\Delta_{23}$ term is relevant in 
eq.~(\ref{4phen}\sep{c}) as far as the LSND experiment is concerned. The LSND
oscillation probability can thus approximately be written as 
$P=A\cdot \sin^2\Delta_{23}$, 
where $A = 4 c_e c_\mu \cos^2\phi_{\mu e}$. 
The LSND results can thus be plotted conveniently in the 
$A$--$|\Delta m^2_{23}|$ plane.  Moreover, the $\nu_e\rightarrow\nu_e$ 
and $\nu_\mu\rightarrow\nu_\mu$ disappearance experiments set upper limits 
both on the amplitudes of $\sin^2\Delta_{23}$ and $\sin^2\Delta_{34}$ on 
the {\em r.h.s} of eqs.~(\ref{4phen}\sep{a,}\sep{b}). We are interested in 
particular in the limits on $4c_e(1-c_e)$ and $4c_\mu(1-c_\mu)$, which 
correspond to two possible ranges for $c_e$ and $c_\mu$, each close to
zero or one.  For $c_e$, however, only the range around zero is allowed since
the other close to one would suppress solar neutrino oscillations in an
unacceptable way. Therefore we have a limit on $c_e$ in the form 
$0\leq c_e\leq a^0_e$ ~\cite{bilenkii:98b}. Concerning the range of 
$c_\mu$, the $\sin^2\Delta_{23}$
term in eq.~(\ref{4phen}\sep{b}) is approximately constant in atmospheric 
neutrino experiments besides being constrained by disappearance experiments 
so that $c_\mu^2\sin^2 2\theta_\mu$ must account for the zenith angle 
dependence of the measured $\nu_\mu$ flux.  This selects out of the two 
possible ranges for $c_\mu$ the interval close to one, that can be written 
as $1-a^0_\mu\leq c_\mu\leq 1$.  The other interval around zero would 
furthermore give a too small amplitude for the $\sin^2\Delta_{23}$ term in 
eq.~(\ref{4phen}\sep{c}) accounting for the LSND signal, since it would 
give $A\leq 4a^0_ea^0_\mu$, which would exclude the LSND result 
(see fig.~\ref{fig:LSND})~\cite{bilenkii:98b}.
Finally, since as said $c^2_\mu\sin^2 2\theta_\mu\simeq \sin^2 2\theta_\mu$
controls the zenith angle dependence of the atmospheric $\nu_\mu$ flux, we also
have $\sin^2 2\theta_\mu\simeq 1$.
The constraint given by the LSND experiment on the parameters $c_e$, $c_\mu$,
$\cos^2\phi_{\mu e}$, namely $4 c_e c_\mu \cos^2\phi_{\mu e} = A$, has not been
used in previous analysis. As we will see, it will play an important role in
the following. 
The solar result does not give any further constraints for this 
discussion since we are in the limit where $\Delta m^2_{12}=0$.

In the following subsection we will use eqs.~(\ref{12},\ref{12par}) and the
parameter constraints discussed above to study CP-violation in long- and
short-baseline experiments.

\subsection{Long- and short-baseline $\nu_\mu\rightarrow\nu_e$}

Let us first identify the allowed parameter range. The viable 
scheme A) of squared mass differences implies of course $|\Delta m^2_{34}|=
\Delta m^2_{\text{ATM}}=(0.3\mbox{--}9)\cdot 10^{-3}\eV^2$ and $|\Delta
m^2_{23}|$ 
is of course the LSND range, 
$|\Delta m^2_{23}| = \Delta m^2_{\text{LSND}} = (0.04\mbox{--}2.8)\eV^2$. 
Furthermore the following constraints  
\globallabel{co}
\begin{align}
& 0 \leq c_e \leq a^0_e \quad 1-a^0_\mu \leq c_\mu \leq 1 & 
& \text{from $\nu_e\rightarrow\nu_e$, $\nu_\mu\rightarrow\nu_\mu$
experiments} \mytag \\
& c_e c_\mu\cos^2\phi_{\mu e}\leq (1-c_e)(1-c_\mu) & & \text{from unitarity}
\mytag \\
& 4 c_e c_\mu\cos^2\phi_{\mu e} = A & & \text{from LSND} \mytag
\end{align}
have to be simultaneously satisfied. It is shown in Appendix B that 
eqs.~(\ref{co}) can be fulfilled if and only if
\begin{equation}
A\leq\min\left(4a^0_\mu(1-a^0_\mu),4a^0_e(1-a^0_e)\right) \simeq
\min\left(4a^0_\mu,4a^0_e\right)~.
\label{coA}
\end{equation}
This gives a restriction on the $|\Delta m^2_{23}|$ range shown in
fig.~\ref{fig:LSND} (the shadowed region is excluded, the function 
$\min(4a^0_\mu,4a^0_e)$ can easily be recovered from the dashed lines).

From eqs.~(\ref{12},\ref{12par}) it is easy 
to see that the CP-asymmetries $a^\CPC_{\nu_\mu \nu_e}$ do not 
depend explicitely on $c_e$, $c_\mu$. Eq.~(\ref{co}\sep{c}) 
introduces however a dependence on $c_e$, $c_\mu$ since 
$\cos^2 \phi_{\mu e}$ is a function of $c_e$, $c_\mu$. It is therefore 
important to know the allowed ranges of $c_e$ and $c_\mu$. 
Eq.~(\ref{co}\sep{a}) alone does not guarantee that it is
possible to fulfill eqs.~(\ref{co}\sep{b,}\sep{c}). 
It turns out that it is possible to find $c_\mu$ and $\phi_{\mu e}$ 
by solving (\ref{co}\sep{b,}\sep{c}) if and only if
\begin{equation}
\label{ran1e}
\frac{A}{4}\simeq\frac{1-\sqrt{1-A}}{2}\leq c_e \leq
\min\left(a^0_e,1-\frac{A}{4a^0_\mu}\right)~.
\end{equation}
This is a non-wide range when eq.~(\ref{coA}) is fulfilled. 
For a given $c_e$ in the range of eq.~(\ref{ran1e}) the possible 
values of $c_\mu$ are those fulfilling simultaneously
\begin{equation}
\label{ran2u}
\frac{A}{4c_e} \leq c_\mu \leq
1-\frac{A}{4(1-c_e)}\qquad\text{and}\qquad c_\mu \geq 
1-a^0_\mu~.
\end{equation}
In this case $A/(4c_ec_\mu)\leq 1$ and $\phi_{\mu e}$ is determined by
\begin{equation}
\cos^2\phi_{\mu e} = \frac{A}{4 c_e c_\mu}~,
\label{cop}
\end{equation}
where $c_\mu=1$ can be used as a good 
approximation\footnote{More details on the statements 
of this paragraph can be found in Appendix B.}.

Finally we have all the necessary ingredients to discuss CP-violation in terms
of the quantity $|a^\CPC_{\nu_\mu \nu_e}/\sin\delta_{\mu e}|$ which
depends mostly only on $|\Delta m^2_{34}|$, $|\Delta m^2_{23}|$, $c_e$, with
the ranges given by eq.~(\ref{Dm2}\sep{a}), eq.~(\ref{coA}) and
eq.~(\ref{ran1e}) respectively. The dependence on $\sin^2 2\theta_\mu$ is in
fact negligible since $\sin^2 2\theta_\mu\simeq 1$ and we can therefore set
$\sin^2 2\theta_\mu= 1$ without affecting our results too much. Analogously, we
set $c_\mu=1$. The dependence of $|a^\CPC_{\nu_\mu \nu_e}/\sin\delta_{\mu e}|$
on $\delta_{\mu e}$ is also negligible in the limit $|\Delta m^2_{34}|\ll
|\Delta m^2_{23}|$.

\begin{figure}
\begin{center}
\epsfig{file=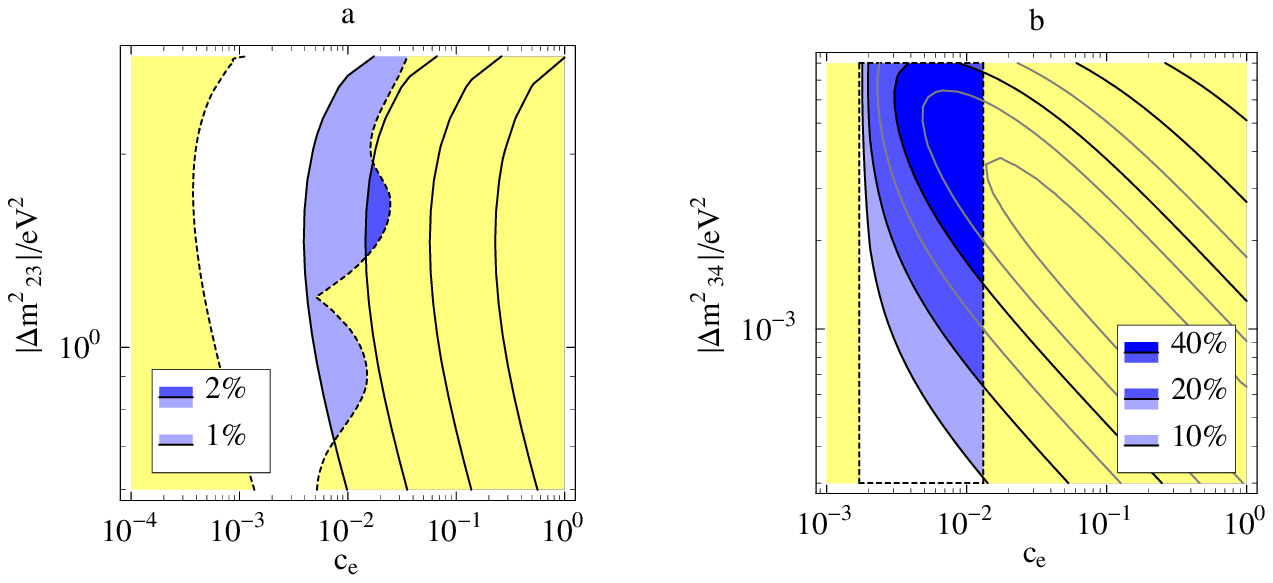,width=0.9\textwidth}
\end{center}
\mycaption{Contour lines for
$|a^\CPC_{\nu_\mu \nu_e}/\sin\delta_{\mu e}|$ in a four neutrino scenario
plotted in the $c_e$--$|\Delta m^2_{23}|$ parameter space of a
short-baseline experiment for $|\Delta m^2_{34}| = 8\cdot
10^{-3}\eV^2$ \textrm{(a)} and in the $c_e$--$|\Delta m^2_{34}|$ plane
of a long-baseline experiment for $|\Delta m^2_{23}| = 1\eV^2$
\textrm{(b)}. Only the allowed $|\Delta m^2|$ ranges are
shown. The shadowed regions are excluded by the
constraints~(\ref{ran1e}).}
\label{fig:asy4me}
\end{figure}

As a consequence of the Bugey limit, the amplitude of the
$\nu_\mu\rightarrow\nu_e$ oscillation in a short-baseline experiment
is necessarily small. One may wonder if such a small oscillation can
be accompanied by relatively large CP-violation as it can happen in 
the three neutrino scenario. To see that this is not the case it is enough 
to study the qualitative features of 
$|a^\CPC_{\nu_\mu \nu_e}/\sin\delta_{\mu e}|$ analytically. 
With the approximations $\sin^2\Delta_{24}\simeq \sin^2\Delta_{23}$ and 
$\sin^2 2\theta_{\mu}=1$ one gets
\begin{equation}
\label{ShAsy}
\fracwithdelims{|}{|}{a^\CPC_{\nu_\mu\nu_e}}{\sin\delta_{\mu e}} \simeq 
\fracwithdelims{|}{|}{4\sin\phi_{\mu e}\cos\phi_{\mu e}
\vev{\sin^2\Delta_{23}\sin\Delta_{34}}}
{4\cos^2\phi_{\mu e}\vev{\sin^2\Delta_{23}}-\cos 2\phi_{\mu e}
\vev{\sin^2\Delta_{34}}}
\end{equation}
and in particular
\begin{equation}
\label{ShAp}
\fracwithdelims{|}{|}{a^\CPC_{\nu_\mu\nu_e}}{\sin\delta_{\mu e}} \simeq 
\left|\tan\phi_{\mu e} 
\frac{\vev{\sin^2\Delta_{23}\sin\Delta_{34}}}{\vev{\sin^2\Delta_{23}}}
\right|
\end{equation}
for short-baseline experiments. Even in the ``CP-optimistic''case with 
$|\Delta m^2_{23}|=8\cdot 10^{-3}\eV^2$
eq.~(\ref{ShAp}) is rather strongly suppressed by $\sin\Delta_{34}$ in a 
short-baseline experiment. Measurable effects could only show up 
if an enhancement by a large $\tan\phi_{\mu e}$ were possible,
which would require via eq.~(\ref{cop}) a rather small value of $A/4c_e$.
The smallest value of $A$ allowed by the LSND plot and the maximal 
value of $c_e$ allowed by eq.~(\ref{ran1e}) do however not allow 
values of $A/c_e$ much smaller than one. CP-violation is consequently
expected to be negligible which is confirmed by fig.~\ref{fig:asy4me}a, 
where the contour lines for $|a^\CPC_{\nu_\mu \nu_e}/\sin\delta|$ 
are plotted in the $c_e$--$|\Delta m^2_{23}|$ plane using the exact 
formulas eqs.~(\ref{12},\ref{12par}) for the oscillation probability  
for an optimistic value of $|\Delta m^2_{34}|=8\cdot 10^{-3}\eV^2$. 
The assumed experimental setup is $L\simeq 0.5\,\text{km}$ and a 
broad distribution for $E$ around $1\GeV$ which looks like MiniBooNE.
Fig.~\ref{fig:asy4me}a shows that the asymmetry does not exceed 3\%,
even for maximal phase.

Long-baseline experiments are of course not $\sin\Delta_{34}$ suppressed and 
the $\sin^2\Delta_{23}$ dependence becomes negligible since it is washed out 
by averaging over the $L/E$ spectrum. Eq.~(\ref{ShAsy}) becomes thus
\begin{equation}
\label{ShAsy2}
\fracwithdelims{|}{|}{a^\CPC_{\nu_\mu\nu_e}}{\sin\delta_{\mu e}} \simeq 
\fracwithdelims{|}{|}{2\sin\phi_{\mu e}\cos\phi_{\mu e}
\vev{\sin\Delta_{34}}}{2\cos^2\phi_{\mu e}-\cos 2\phi_{\mu
e}\vev{\sin^2\Delta_{34}}}~.
\end{equation}
CP-violation in this case is not suppressed at all. This can be seen directly
from fig.~\ref{fig:asy4me}b, where contour lines for $|a^\CPC_{\nu_\mu
  \nu_e}/\sin\delta_{\mu e}|$ are plotted in the $c_e$--$|\Delta m^2_{34}|$
plane for $|\Delta m^2_{23}|=1\eV^2$ in a ``MINOS-like'' long-baseline
experiment like described in the previous section. The unshadowed rectangular
window in fig.~\ref{fig:asy4me}b represents the values of $c_e$ which are
allowed by the Bugey experiment and the unitarity constraint.  The CP-asymmetry
in the allowed region can reach 60\% for a maximal CP-violating phase. This
leads to the important question whether the allowed region could be reached by
a long-baseline experiment despite the strong constraint from Bugey.  The
answer depends on the value of $|\Delta m^2_{23}|$ and $\delta_{\mu e}$ but in
many cases is yes.  
A definitive answer would require a plot of the sensitivity of a long-baseline
experiment (here we consider MINOS again) in the allowed region of the
parameter space. The published sensitivity plots are however for the parameter
space $\{A,\Delta m^2\}$ of a simple two neutrino oscillation in which the
transition probability is given by $P=A\sin^2\Delta m^2 L/(4E)$. The
$\nu_\mu\rightarrow \nu_e$ transition probability cannot be reduced to that
form in our case being rather
\begin{equation}
\label{TrPr}
P(\nu_\mu\rightarrow\nu_e) \simeq A/2 +c_e(-\cos 2\phi_{\mu e}
\vev{\sin^2\Delta_{34}}\pm\sin 2\phi_{\mu e}\sin\delta_{\mu e}
\vev{\sin\Delta_{34}})~,
\end{equation}
where the sign of the $\vev{\sin\Delta_{34}}$ term depends on the sign of 
$\sin 2\theta_\mu$ and can be reabsorbed in the definition of $\delta_{\mu
e}$. $A$ is the LSND amplitude given in fig.~\ref{fig:LSND} and
eq.~(\ref{co}\sep{c}) has been used. The sensitivity in the 
$c_e$--$|\Delta m^2_{34}|$ plane depends therefore on $\sin\delta_{\mu e}$,
which controls the interference in eq.~(\ref{TrPr}), and on $A$ 
(and in turn on $|\Delta m^2_{23}|$) mainly through the $A/2$ term. 
Clearly larger values of $A$ are preferred and it turns out that 
the sensitivity is larger for positive values of $\sin\delta$. For example 
the sensitivity obtained for $\sin\delta=1$ and $|\Delta m^2_{23}|=1\eV^2$
allows to reach most of parameter space. Values of $A$ which are in better
agreement the KARMEN experiment give less sensitivity and the smallest possible
value of $A$ allowed by LSND at 99\% CL would give a transition probability too
small to be measured. Nevertheless a long-baseline $\nu_\mu\rightarrow\nu_e$
experiment along the line discussed has in a four neutrino scenario a good
chance to observe oscillation and this oscillation would most likely contain a
sizable or even large CP-violating part.

\subsection{Long- and short-baseline $\nu_\mu\rightarrow\nu_\tau$} 

\begin{figure}[htb]
\begin{center} 
\epsfig{file=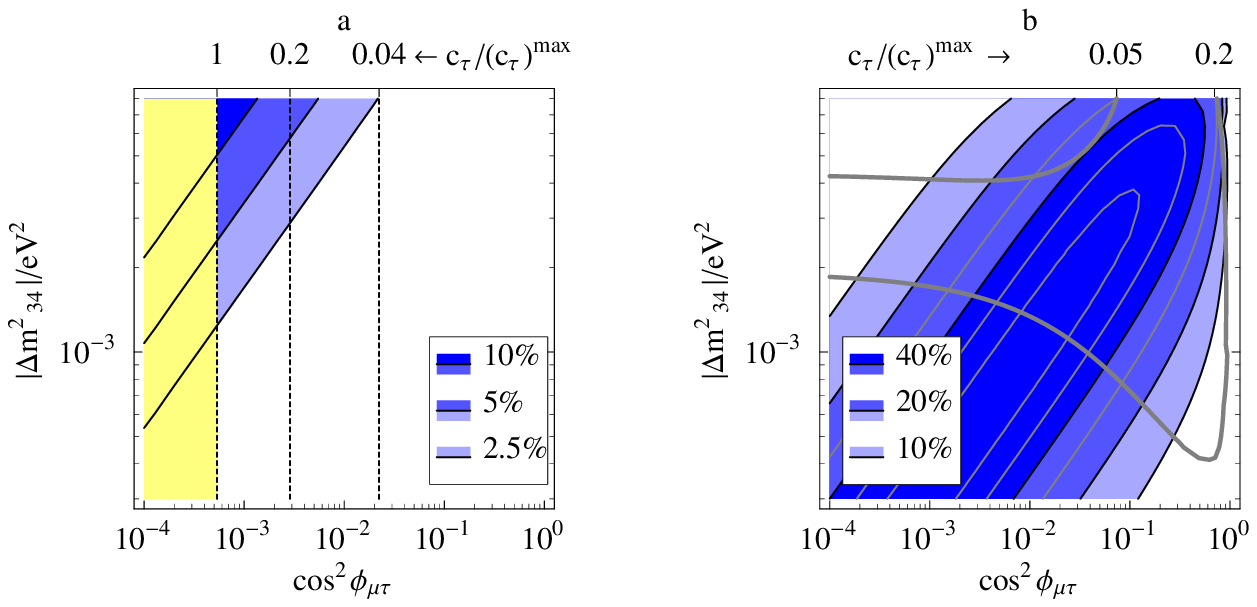,width=0.9\textwidth} 
\end{center} 
\mycaption{Contour lines for $|a^\CPC_{\nu_\mu\nu_\tau}/\sin\delta_{\mu\tau}|$
in the $\cos^2\phi_{\mu\tau}$--$|\Delta m^2_{34}|$ plane for $|\Delta 
m^2_{23}|=1\eV^2$ (a) for the short-baseline experiment and (b) for 
the long-baseline experiment discussed in the text. The accessible 
parameter space of such an experiment depends on $c_{\tau}$. The 
non-shadowed region shows the optimal case of $c_{\tau} = c_{\tau}^{max}$ 
which covers the whole plot in (b). The dashed lines show the sensitivity 
in less favourable cases with $c_{\tau} < c_{\tau}^{max}$ (see labels 
above figure and text). Everything to the right of the dashed lines in 
(a) can be explored by the discussed experiment whereas in 
(b) the whole area from the upper-left corner to the dashed lines lies 
within the sensitivity of the experiment.}
\label{fig:asy4mu} 
\end{figure}

Unlike what we have seen in the three neutrino case, here the
$\nu_\mu\rightarrow\nu_\tau$ channel turns out to be very interesting. 
From the previous subsection it is immediately clear that the
relevant parameters are now $|\Delta m^2_{23}|$, $|\Delta m^2_{34}|$,
$\theta_\mu$, $c_\mu$, $c_\tau$, $\phi_{\mu\tau}$ and $\delta_{\mu\tau}$. The
parameters $|\Delta m^2_{23}|$, $|\Delta m^2_{34}|$, $\theta_\mu$ and $c_\mu$
are constraint as before. The unitarity constraints eqs.~(\ref{4uni}) for the
$\mu\tau$/34 sub-matrix give now additionally the conditions 
\globallabel{cou}
\begin{gather} 
0\leq c_\tau \leq 1 \mytag ~,\\ 
c_\mu c_\tau\cos^2\phi_{\mu\tau}\leq (1-c_\mu)(1-c_\tau)~, \mytag 
\end{gather} 
while $c_\mu$ is already known to be in the range  
\begin{equation} 
\label{ran1u} 
\max(1-a^0_\mu,\frac{A}{4a^0_e}) \equiv c_\mu^{\text{min}} \leq c_\mu 
\leq \frac{1+\sqrt{1-A}}{2}~. 
\end{equation} 
In addition to the transition probabilities already given in 
eqs.~(\ref{4phen}) there is now the $\nu_\mu\rightarrow\nu_\tau$ channel 
\begin{multline} 
\label{uPro} 
P(\nu_\mu\rightarrow\nu_\tau) \simeq 4 c_\mu c_\tau \cos^2\phi_{\mu\tau} 
\sin^2\Delta_{23} +2 c_\mu c_\tau 
\sin 2\theta_\mu\sin 2\phi_{\mu\tau}\sin\delta_{\mu\tau} 
\sin^2\Delta_{23}\sin\Delta_{34} \\ 
-c_\mu c_\tau \sin 2\theta_\mu (\cos 2\phi_{\mu\tau}\sin 2\theta_\mu 
+\sin 2\phi_{\mu\tau}\cos 2\theta_\mu\cos\delta_{\mu\tau}) 
\sin^2\Delta_{34}~.
\end{multline} 
Let us consider first a short-baseline experiment.  Among the oscillating 
terms only $\sin^2\Delta_{23}$ can develop and contribute to this transition 
probability. The {\em r.h.s.} of eq.~(\ref{uPro}) is therefore to a very 
good approximation given by the first term which is furthermore suppressed 
by the unitarity constraints on $c_\mu$ and $c_\tau$. This suppression 
turns out to be much less effective in the CP-violating part, whose 
relative importance grows therefore when the total probability gets smaller, 
very much like in the three neutrino $\nu_\mu\rightarrow\nu_e$ case. 
One obtains for the short-baseline case
\begin{equation} 
\label{ShApu} 
\fracwithdelims{|}{|}{a^\CPC_{\nu_\mu\nu_\tau}}{\sin\delta_{\mu\tau}} \simeq  
\left|\tan\phi_{\mu\tau} 
\frac{\vev{\sin^2\Delta_{23}\sin\Delta_{34}}}{\vev{\sin^2\Delta_{23}}} 
\right|~,
\end{equation}
and we can see that the $\sin\Delta_{34}$ suppression can be compensated by 
large values of $\tan\phi_{\mu\tau}$.  In fig.~\ref{fig:asy4mu}a we show 
contour lines for $|a^\CPC_{\nu_\mu\nu_\tau}/\sin\delta_{\mu\tau}|$
in the $\cos^2\phi_{\mu\tau}$--$|\Delta m^2_{34}|$ plane for 
$|\Delta m^2_{23}|=1\eV^2$ for the short baseline experiment described 
in the previous subsection. The maximum possible sensitivity of such 
an experiment is within the non-shadowed region. The sensitivity in 
less favourable cases is also shown. The precise way how the sensitivity 
was here obtained and the meaning of the ``less favourable cases'' is 
explained in greater detail in the following long-baseline case. 
Contrarily to what happened in the three neutrino $\nu_\mu\rightarrow\nu_e$ 
case, in which the necessary enhancement was obtained within the sensitivity 
of the experiment, the necessary enhancement is here outside the reach of 
an experiment like MiniBooNE. Note however that an enhancement by a 
factor $\sim 10$ in $\vev{L/E}\sim 0.5$ of the assumed short-baseline
experiment would be enough to test CP-violation, which can be seen by 
multiplying the asymmetry in fig.~\ref{fig:asy4mu}a by $\vev{L/E}/0.5$. 
It is therefore not necessary to go a long-baseline one with $\vev{L/E}\sim
100$ to test CP-violation.

Finally we can also discuss in the four neutrino case CP-violation in a
long-baseline experiment like that considered in the three neutrino case. The
corresponding contour lines for
$|a^\CPC_{\nu_\mu\nu_\tau}/\sin\delta_{\mu\tau}|$ are shown in
fig.~\ref{fig:asy4mu}b as before in the $\cos^2\phi_{\mu\tau}$--$|\Delta
m^2_{34}|$ plane for fixed $|\Delta m^2_{23}|=1\eV^2$. The precise value of
$|\Delta m^2_{23}|$ is anyway irrelevant since $\sin^2\Delta_{23}$ is 
averaged to $1/2$ in this case.  The asymmetry is in this case also 
completely independent of $c_\mu$, $c_\tau$ and the dependence on 
$\theta_\mu$ is negligible as before.
What matters is again the region of parameter space accessible to the
experiment, which depends on $c_\tau$ in the way given by eq.~(\ref{uPro}). 
The range of $c_\tau$, which is determined by eqs.~(\ref{cou}) in terms of
$\cos^2\phi_{\mu\tau}$, is therefore crucial and one finds
\begin{equation} 
\label{boh} 
0\leq c_\tau \leq 
\frac{1-c_\mu^{\text{min}}}{1-c_\mu^{\text{min}}\sin^2\phi_{\mu\tau}}~. 
\end{equation} 
The sensitivities of fig.~\ref{fig:asy4mu}b depend therefore on the chosen
value of $c_\tau$ and become maximal for the largest possible value
$c_\tau^{\text{max}}\equiv
\frac{1-c_\mu^{\text{min}}}{1-c_\mu^{\text{min}}\sin^2\phi_{\mu\tau}}$.  For
this maximal value $c_\tau=c_\tau^{\text{max}}$ one finds that the experimental
sensitivity covers the complete coordinate range of fig.~\ref{fig:asy4mu}b and
only rather small values of $c_\tau$ make some of the relevant parts of the
parameter space unaccessible.  We show in fig.~\ref{fig:asy4mu}b the
sensitivity corresponding to $c_\tau = 0.2\, c_\tau^{\text{max}}$ and $c_\tau
= 0.05\, c_\tau^{\text{max}}$. The region to the right of the 20\% (5\%)
sensitivity line are excluded if $c_\tau$ is less than 20\% (5\%) of its
maximum value in that region. One can see that $c_\tau = 0.2\,
c_\tau^{\text{max}}$ still allows to reach the interesting region whereas
$c_\tau = 0.05\, c_\tau^{\text{max}}$ excludes it.
Here, as in the $\nu_\mu\rightarrow\nu_e$ case, the sensitivity depends on the
sign of $\sin 2\theta_\mu$ and on the CP-violating phases, that control the
interference between the CP-violating and CP-conserving parts of the
transition probabilities. The lines plotted correspond to $\sin 2\theta_\mu
=1$, $\sin\delta_{\mu \tau}=1$, that give a better sensitivity.
The results of this subsection show that $\nu_\mu\rightarrow\nu_\tau$ is a good
place to look for CP-violation in four neutrino scenarios even with
intermediate-baseline experiments. Moreover, if there are four neutrinos and if
a signal is observed, CP-violation should be taken into account in an analysis.


\section{Discussion and Conclusions \label{sec:concl}}

We studied in this paper CP-violation in neutrino oscillation and 
discussed the sensitivity for such effects of current and future 
experiments. 
As a measure of CP-violation, we considered asymmetries between 
CP-conjugated transition probabilities. These asymmetries are a 
measure of the relative strength of CP-violating effects. They are 
therefore very useful in studying how much CP-violation would affect 
the measurement of oscillation in one of the two CP-conjugated channels.
The contribution from CP-violation has to be separated however from 
the experimental asymmetries between the two channels, which we
assumed to be small or under control. 

The fact that it is not possible to accommodate the 
claimed three independent oscillation signals in scenarios with 
three neutrinos led us to a twofold strategy. The first scenario was 
to leave out one evidence for oscillation and to analyze the case 
of three neutrinos. Since the LSND evidence is almost in contradiction 
with KARMEN it is considered most controversial and we omitted 
it therefore in the three neutrino case. The second case which
was studied includes LSND in a four neutrino scenario. In both cases
all further existing exclusion limits were taken into account. The two 
scenarios lead to quite different results with different sizes of 
CP-violating effects. In all cases we present results for maximal 
CP-phase, which can be easily rescaled to an arbitrary value.

For three neutrinos CP-violating effects are drastically suppressed 
by small angles or by extremely small $\Delta m^2$ in the case of the 
small mixing angle MSW-solution and for the vacuum solution. 
The large mixing angle MSW-solution allows however sizable CP-violating 
effects in long-baseline experiments, while there is no effect in 
short-baseline experiments. In the $\mu\rightarrow e$ channel we found 
in long-baseline experiments for maximal CP-phase $\delta$ effects up 
to 40\%, while we found only small effects in the $\mu\rightarrow \tau$ 
channel. The observation of CP-violation would also allow to distinguish 
between the different solar solutions and can even further restrict the 
parameter space for $|\Delta m_{12}^2|$. 
Including matter effects in these long baseline experiments we found
for the considered setup moderate corrections compared to the case 
without matter. Depending on the sign of the squared mass difference,
the total intrinsic CP-asymmetry in $\mu\rightarrow e$ oscillation 
can be enhanced up to 50\%. In $\mu\rightarrow \tau$ oscillation 
matter effects dominate the asymmetry since the intrinsic CP-violation
is very small and therefore anyway uninteresting. 

In order to include the LSND result we studied also the case of four 
neutrinos where many more parameters exist in principle. The solar 
$\Delta m^2$-value is in this case unimportant since CP-violating effects 
can be generated by the larger $\Delta m^2$ responsible for the 
atmospheric neutrino oscillations and for the LSND measurement. For 
experiments which are only sensitive to $\Delta m^2 \geq 10^{-3}~eV^2$, 
we could make the approximation $\Delta m_{12}^2 =0$ which reduces the 
number of relevant parameters drastically and allows a study of the 
available parameter space. The CP-violating effects are now in the case 
of four neutrinos potentially larger and we considered therefore also 
short-baseline experiments. Altogether we find in the four neutrino
scenario for maximal CP-phase the following effects:
In short-baseline $\mu\rightarrow e$ experiments less than 2~\%,
in short-baseline $\mu\rightarrow \tau$ up to 10~\%  and
in long-baseline $\mu\rightarrow e$ 
as well as  $\mu\rightarrow \tau$ experiments up to 60~\%.
The effects are not very big in the considered MiniBooNE-like 
short-baseline setup, but we want to point out, that a modest improvement 
of $\vev{L/E}\simeq 0.5$ by a factor 10 would be enough to test 
CP-violation in the $\nu_\mu\rightarrow\nu_\tau$ channel.

We did not consider cases with more than four neutrino mass eigenstates. 
It should however be clear from the current analysis that large CP-violating
effects could easily be involved in that case in current and/or future 
neutrino oscillation experiments.

In summary we found that CP-violating effects can be surprisingly large 
in some future neutrino oscillation experiments and such effects should 
therefore be included in the analysis. Besides the obvious general 
interest for a determination of CP-violation connected to the theoretical 
questions on physics beyond the Standard Model and the potential role which 
CP-violation could play in lepto-- and baryogenesis, there are further 
reasons why CP-violation should be taken into account. The main point
is that the omission of CP-violation can spoil a two or three neutrino 
analysis that does not take it into account. Moreover, if an asymmetry 
between CP-conjugated transitions were measured and the presence of 
light sterile neutrinos would be excluded, it would discriminate between 
the different solar solutions and set lower bounds on the solar $\Delta m^2$.

\vspace*{.5cm} Acknowledgments: We are grateful to E.~Akhmedov,
S.~Bilenky and M.B.~Gavela for useful comments and discussions. 
AR wishes to thank the Institutes T30d and T31 at the Physics 
Department of the Technical University of Munich for their warm 
hospitality.


\newpage
\appendix

\section*{Appendix A}

A non-standard solar analysis allows in principle the possibility 
that $|\Delta m^2_{12}|$ falls in the atmospheric range, but still 
consistent with the Chooz constraint. For this case one may wonder 
whether 
\begin{itemize}
\item[{\bf ~i)}] $|\Delta m^2_{12}|$ could be responsible for 
both the solar evidences (with Chlorine excluded) and the atmospheric 
results with $|\Delta m^2_{23}|$ above the atmospheric range (to 
account for cosmological requirements on neutrino masses)  
\end{itemize}
and whether
\begin{itemize}
\item[{\bf ii)}]  $|\Delta m^2_{23}|$ could explain the LSND signal.  
\end{itemize}
The answers to these two questions are that, independently of {\bf ii)}, 
{\bf i)} is strongly disfavoured and that {\bf ii)} is not possible. 
{\bf i)} is disfavoured independently of the possibility of explaining 
LSND for two reasons. First of all the probability that $|\Delta m^2_{12}|$ 
is in the part of the atmospheric range which is not excluded by the 
Chooz constraint is small. This, together with the solar constraints
which exclude large $|U_{e3}|^2$ gives $|U_{e3}|^2 \lesssim 0.025$.
The survival probability for solar 
neutrino state $\nu_2$ is therefore
$P(\nu_e\rightarrow\nu_e) = 1 - \sin^2 2\theta_{12} \sin^2\Delta_{12}$ 
with $\sin^2 2\theta_{12}\simeq 1$ 
which has to explain the solar data (with Chlorine 
excluded). This is on the other hand in this scenario also the survival 
probability for atmospheric $\nu_e$ appearing in eq.~(\ref{RmR}\sep{b}). 
The atmospheric fit would in this case therefore be very bad, since the 
the zenith angle variation of $R_e$ in eqs.~(\ref{RmR}) would be larger 
than for $R_\mu$.

LSND cannot be explained in any case, even if the two points above are 
ignored. The LSND oscillation probability is for this scenario 
\begin{equation}
P(\bar\nu_\mu\rightarrow\bar\nu_e) = 4|U_{e3}U_{\mu 3}|^2 \sin^2\Delta_{23}~.
\label{PLSND}
\end{equation}
The mixing $|U_{e3}|^2$ is known to be small with the upper limit 
given by Bugey. The mixing $|U_{\mu 3}|^2$ is also small because 
$4|U_{\mu 1}U_{\mu 2}|^2 \leq (1-|U_{\mu 3}|^2)^2 $ has to be 
maximal in order to explain the oscillation of $R_\mu$  in 
eq.~(\ref{RmR}\sep{a}) and because of the experimental limit given 
by CDHS and CCFR. Combining these two limits into a single limit 
for $4|U_{e3}U_{\mu 3}|^2$ one finds that the probability in
eq.~(\ref{PLSND}) is too small to explain the LSND signal~\cite{bilenkii:98a}. 

\section*{Appendix B} 
\label{app}

In this appendix we prove the statements concerning the range of
$c_e$, $c_\mu$, $\phi_{\mu e}$ satisfying eqs.~(\ref{co}).
First we prove that eq.~(\ref{coA}) is sufficient for the existence of
a solution to these equations. Therefore it is enough to check that
$c_e=c^{\text{min}}$, $c_\mu=c^{\text{max}}$, $\phi_{\mu e}=0$ 
are a solution of  
eqs.~(\ref{co}).  
Then we prove that eq.~(\ref{coA}) is also a necessary condition.
From eqs.~(\ref{co}\sep{b,}\sep{c}) follows $4c_e c_\mu\geq A$
and $4(1-c_e)(1-c_\mu) \geq A$ and therefore $c_e \geq A/(4c_\mu)$,
$c_\mu \leq 1-A/(4(1-c_e))$. In particular we have $4c_e^2-4c_e+A \leq 0$ 
and $4c_\mu^2-4c_\mu+A \leq 0$ so that, together with the previous 
relations, we obtain
\[
c_e \geq c^{\text{min}} = \frac{1-\sqrt{1-A}}{2} \simeq \frac{A}{4} ~;\qquad 
c_\mu \leq c^{\text{max}} = \frac{1+\sqrt{1-A}}{2} \simeq
1-\frac{A}{4}~. 
\]
But in order to satisfy eq.~(\ref{co}\sep{a}) one must have
$c^{\text{min}} \leq a^0_e$ and
$c^{\text{max}} \geq 1 - a^0_\mu$, namely
\[
A\leq 4a^0_e(1-a^0_e) \quad\text{and}\quad A\leq 4a^0_\mu(1-a^0_\mu)~.
\]

Let us now prove eq.~(\ref{ran1e}). To begin, let us consider a given
value of $c_e$. Eqs.~(\ref{co}\sep{b,}\sep{c}) give
\begin{equation}
\label{ranm}
\frac{A}{4 c_e} \leq c_\mu \leq 1-\frac{A}{4(1-c_e)}~,
\end{equation}
where a solution exists when eq.~(\ref{coA}) holds. 
Since $1-a^0_\mu \leq c_\mu \leq 1$, $1-A/(4(1-c_e))$ must be equal to or
larger than $1-a^0_\mu$ in order to obtain a finite range for $c_\mu$. One 
finds thus
\[
c_e\leq 1-\frac{A}{4a^0_\mu}~,
\]
which, together with the previous bounds, gives eq.~(\ref{ran1e}). 
To see that eq.~(\ref{ran1e}) is not an empty interval we have to check that
$c^{\text{min}}_e\leq 1-A/(4a^0_\mu)$. This is a consequence of
$A=(1-c^{\text{min}}_e) (1-c^{\text{max}}_\mu)$ and
$c^{\text{max}}_\mu\geq 1-a^0_\mu$.  
The range of $c_\mu$ for a given value of $c_e$, eq.~(\ref{ran2u}), then
follows from what we saw.


\newpage

\bibliographystyle{phaip}

\begin{thebibliography}{1}

\bibitem{others}
V. Barger, Yuan-Ben Dai, K. Whisnant and Bing-Lin Young,
\newblock Neutrino Mixing, CP/T Violation and Textures in Four-Neutrino Models,
\newblock hep-ph/9901388;\\
K.R. Schubert,
\newblock May We Expect CP- and T-Violating Effects in Neutrino Oscillations?,
\newblock hep-ph/9902215;\\
J. Arafune, M. Koike and J. Sato, \newblock Phys.~Rev.~{\bf D56} (1997) 3093;\\
M. Tanimoto, \newblock Phys.~Lett.~{\bf B435} (1998) 373;\\
H. Minakata and H. Nunokawa, \newblock Phys.~Rev.~{\bf D57} (1998) 4403.

\bibitem{gavela98} A.De~R\'ujula, M.B.~Gavela and P.~Hern\'andez,
\newblock Neutrino Oscillation Physics with a Neutrino Factory,
\newblock hep-ph/9811390.

\bibitem{BAHSSM98} 
J.N. Bahcall, S. Basu and M.H. Pinsonneault, 
\newblock Phys.~Lett.~{\bf B433} (1998) 1.

\bibitem{bahcall:98a}
J.~N. Bahcall, P.~I. Krastev and A.~Y. Smirnov,
\newblock Phys.~Rev.~{\bf D58} (1998) 096016.

\bibitem{SK:98b}
Super-Kamiokande Collaboration,
\newblock Phys.~Rev.~Lett.~{\bf 82} (1999) 1810.

\bibitem{BHSSW}
R.~Barbieri, L.~J. Hall, D.~Smith, A.~Strumia and N.~Weiner,
\newblock Journal of High Energy Physics, JHEP 9812 (1998) 017. 

\bibitem{SK98}
Super-Kamiokande Collaboration, T.~Kajita, in Neutrino 98,
Proceeding of the XVIII International Conference on Neutrino
Physics and Astrophysics, Takayama, Japan, 4-9 June 1998, edited
by Y.Suzuki and Y. Totsuka.

\bibitem{LSND98}
C. Athanassopoulos et al., LSND Coll., 
Phys.~Rev.~{\bf C58} (1998) 2489.

\bibitem{CHOOZ97}
Chooz Collaboration, Phys.~Lett.~{\bf B420} (1998) 397.

\bibitem{CHORUS98}
CHORUS Collaboration, New Results on the $\nu_\mu-\nu_\tau$ Oscillation Search
with the CHORUS Detector, hep-ex/9807024.

\bibitem{bugey}
B.~Achkare et al., Nucl.~Phys.~{\bf B434} (1995) 503.   

\bibitem{CDHS}
F.~Dydak et al., Phys.~Lett.~{\bf B134} (1984) 281.

\bibitem{CCFR}
I.~E.~Stockdale et al., Phys.~Rev.~Lett.~{\bf 52} (1984) 1384.

\bibitem{barger:98a}
V.~Barger, T.~J. Weiler and K.~Whisnant,
\newblock Phys.~Lett.~{\bf B440} (1998) 1.

\bibitem{SK:98a}
{Super-Kamiokande Collaboration},
\newblock Phys.~Rev.~Lett.~{\bf 81} (1998) 1562.

\bibitem{bilenkii:98b}
S.~M. Bilenky, C.~Giunti and W.~Grimus,
\newblock Eur.~Phys.~J.~{\bf C1} (1998) 247.

\bibitem{bilenkii:98a}
S.~M. Bilenky, C.~Giunti and W.~Grimus,
\newblock Phys.~Rev.~{\bf D58} (1998) 033001.

\bibitem{minos} 
MINOS Technical Design Report,\\
http://www.hep.anl.gov/ndk/hypertext/minos\_tdr.html

\end{thebibliography}

\end{document}